%% file: main.tex
\documentclass[9pt]{article}
\def\pagenum{1}

\pdfoutput=1

\input{files/configs}

\usepackage[compress]{natbib}
\usepackage{graphicx}
\usepackage{float}
\usepackage{indentfirst} 
\usepackage{bigstrut,multirow,rotating,booktabs}%表格插件
\usepackage{lineno}
\usepackage{titlesec}
\usepackage{hyperref} %生成引用链接，注：该宏包可能与其他宏包冲突，故放在所有引用的宏包之后
\usepackage{cleveref} %实现图片和表格、公式的引用，cleveref包必须放在hyperref包之后

\hypersetup{colorlinks = true, %将链接文字带颜色
	    linkcolor = blue, %图表引用颜色设置为蓝色
	    urlcolor = blue,%网页链接为蓝色
        citecolor = blue} %文献引用颜色设置为蓝色

\usepackage{etoolbox}
\BeforeBeginEnvironment{tabular}{\footnotesize}

\usepackage{enumitem}
\setitemize[1]{itemsep=0pt,partopsep=0pt,parsep=\parskip,topsep=5pt}

\usepackage{authblk}

\crefname{table}{Table}{Table}
\crefname{figure}{Figure}{Figure}

\def\thetitle{{Operators' cognitive performance under extreme hot-humid exposure and its physiological-psychological mechanism based on ECG, fNIRS, and Eye Tracking}} 
\def\authorOne{\authorfont{Yan Zhang}}
\def\authorTwo{\authorfont{Ming Jia}}
\def\authorThree{\authorfont{Meng Li}}
\def\authorFour{\authorfont{JianYu Wang}}
\def\authorFive{\authorfont{XiangMin Hu}}
\def\authorSix{\authorfont{ZhiHui Xu}}
\def\authorSeven{\authorfont{Tao Chen}}

\def\institutionOne{\subauthorfont{State Key Laboratory of Nuclear Power Safety Technology and Equipment, China Nuclear Power Engineering Co., Ltd., Shenzhen, Guangdong, 518172, China}}
\def\institutionTwo{\subauthorfont{School of Safety Science, Tsinghua University, Beijing 100084,  P.R. China}}
\def\institutionThree{\subauthorfont{Anhui Province Key Laboratory of Human Safety, Hefei Anhui 230601, China}}
\def\institutionFour{\subauthorfont{Beijing Key Laboratory of Comprehensive Emergency Response Science, Beijing, China}}

\begin{document}
% ----------------------------- %

\interfootnotelinepenalty=100000

% ----------------------------- %
%  Header: Title, author
% ----------------------------- %
\title{\TitleFont{\thetitle}}
\author[b]{\authorOne}
\author[a]{\authorTwo}
\author[b]{\authorThree}
\author[b]{\authorFour}
\author[b]{\authorFive}
\author[a,$^*$]{\authorSix}
\author[b,c,d,\footnote{Corresponding author. \\ E-mail addresses: xuzhihui@hrbeu.edu.cn (Z.H. Xu), chentao.a@tsinghua.edu.cn (T. Chen).}]{\authorSeven} 

\affil[a]{\institutionOne}
\affil[b]{\institutionTwo}
\affil[c]{\institutionThree}
\affil[d]{\institutionFour}

\vspace{-1mm}
\date{}                                                 
\maketitle

%\linenumbers
% ----------------------------- %
% Abstract
% ----------------------------- %
\input{files/01abstract}

\thispagestyle{fancy}

% ----------------------------- %
% Sections
% ----------------------------- %
\input{files/02introduction}

\input{files/03method}
\input{files/04result}
\input{files/05discussion}
\input{files/06conclusion}
\input{files/07acknowledge}

%\nolinenumbers

% ----------------------------- %
% References:  BibTeX
% ----------------------------- %
\bibliographystyle{elsarticle-harv}
\bibliography{main.bib}

\end{document}

%% file: files/configs.tex
\usepackage{times}                                     
\usepackage{amsfonts,amsmath,amssymb, amsthm}

\usepackage{latexsym}
\usepackage[a4paper]{geometry}

\usepackage{fancyhdr}
\usepackage{placeins}                           
\usepackage{xcolor}
\usepackage[utf8]{inputenc} % Character encoding
\usepackage{graphicx}%
\usepackage{tabularx}
\usepackage{array}
\newcolumntype{x}[1]{%
>{\raggedleft\hspace{0pt}}p{#1}}%

\usepackage{booktabs}

\usepackage{comment}

\usepackage[labelfont=bf,labelsep=period, skip=5pt]{caption}

\usepackage{pgf,tikz}
\usepackage{mathrsfs}
\usetikzlibrary{arrows}

\usepackage{pgfplots}
\pgfplotsset{width=7cm,compat=1.10}

\usepackage{caption}
\usepackage{subcaption}
\usepackage{float}

\usepackage{outlines}
\usepackage{enumitem}
\setenumerate[1]{label=\arabic*.}
\setenumerate[2]{label=(\alph*).}
\setenumerate[3]{label=\roman*.}
\setenumerate[4]{label=\alph*.}

\RequirePackage{color}
\definecolor{cej@color}{rgb}{0.5,0.37109375,0.59765625}
\definecolor{abs}{rgb}{0.251,0.541,0.72.9}
\definecolor{ffqqqq}{rgb}{1,0,0}
\definecolor{qqqqff}{rgb}{0,0,1}
\definecolor{cqcqcq}{rgb}{0.75,0.75,0.75}
\definecolor{ttqqqq}{rgb}{0.2,0,0}
\definecolor{ffxfqq}{rgb}{1,0.5,0}
\definecolor{uuuuuu}{rgb}{0.27,0.27,0.27}
\definecolor{ffqqff}{rgb}{1,0,1}
\definecolor{xfqqff}{rgb}{0.5,0,1}
\definecolor{xdxdff}{rgb}{0.49,0.49,1}
\definecolor{red}{rgb}{1,0,0}
\definecolor{blue}{rgb}{0,0,1}
\definecolor{ccqqqq}{rgb}{0.8,0,0}
\definecolor{fftttt}{rgb}{1,0.2,0.2}
\definecolor{qqwuqq}{rgb}{0,0.39,0}
\definecolor{qqqqff}{rgb}{0,0,1}
\definecolor{wrwrwr}{rgb}{0.3803921568627451,0.3803921568627451,0.3803921568627451}
\definecolor{sexdts}{rgb}{0.1803921568627451,0.49019607843137253,0.19607843137254902}
\definecolor{rvwvcq}{rgb}{0.08235294117647059,0.396078431372549,0.7529411764705882}
\definecolor{ccqqqq}{rgb}{0.8,0,0}
\definecolor{qqqqff}{rgb}{0,0,1}
\definecolor{wwwwww}{rgb}{0.4,0.4,0.4}
\definecolor{qqwuqq}{rgb}{0,0.39215686274509803,0}
\definecolor{uuuuuu}{rgb}{0.266,0.266,0.266}

\usepackage{titlesec}
\titleformat{\section}[block]{\normalsize\scshape\bfseries{\arabic{section}.}}{}{1em}{}
\titleformat{\section}[block]{\normalsize\scshape\bfseries}{\thesection}{1em}{}
\titleformat*{\subsection}{\normalsize\scshape\itshape\bfseries}
\titleformat*{\subsubsection}{\small\scshape\itshape}
\usepackage{framed}
\usepackage[colorlinks,allcolors=black]{hyperref}
\newcounter{ggbFirstpage}
\newcommand*{\TitleFont}{%
      \usefont{\encodingdefault}{\rmdefault}{}{n}%
      \fontsize{16}{18}%
      \selectfont}

\def\authorfont{\normalfont\fontsize{12}{14}\selectfont\centering}
\def\subauthorfont{\normalfont\fontsize{10}{12}\selectfont\centering}
\definecolor{wrwrwr}{rgb}{0.3803921568627451,0.3803921568627451,0.3803921568627451}
\definecolor{sexdts}{rgb}{0.1803921568627451,0.49019607843137253,0.19607843137254902}
\definecolor{rvwvcq}{rgb}{0.08235294117647059,0.396078431372549,0.7529411764705882}

%% file: files/01abstract.tex
% abstract
%
\bgroup
\color{abs}
\hrule
\egroup

%
%%  DO NOT EDIT  LINES ABOVE                     %

\begin{abstract}
Operators' cognitive and executive functions are impaired significantly under extreme heat stress, potentially resulting in more severe secondary disasters. This research investigated the impact of elevated temperature and humidity (25℃ 60\%RH, 30℃ 70\%RH, 35℃ 80\%RH, 40℃ 90\%RH) on the cognitive functions and performance of operators. Meanwhile, we explored the psychological-physiological mechanism underlying the change in performance by electrocardiogram (ECG), functional near-infrared spectroscopy (fNIRS), and eye tracking physiologically. Psychological aspects such as situation awareness, workload, and working memory were assessed. Eventually, we verified and extended the maximal adaptability model to the extreme condition. Unexpectedly, a temporary improvement in simple reaction tasks but rapid impairment in advanced cognitive functions (i.e. situation awareness, communication, working memory) was obtained above 35℃ WBGT. The best performance in a suitable environment was due to more effective activation in the prefrontal cortex (PFC). With temperature increasing, more mistakes occurred and comprehension was impaired due to drowsiness and lower arousal levels, according to evidence of compensatory effect in fNIRS. In the extreme environment, the enhanced PFC cooperation with higher functional connectivity resulted in a temporary improvement, while depressed activation in PFC, heavy physical load, and poor regulation of the cardiovascular system restricted it. Our results provide a more detailed study of the process of operators' performance and cognitive functions when encountering increasing heat stress, as well as its underlying mechanisms from a neuroergonomics perspective. This can contribute to a better understanding of the interaction between operators' performance and workplace conditions, and help to achieve a more reliable human-centered production system in the promising era of Industry 5.0.


\end{abstract}

\bgroup
\color{abs}
\hrule
\egroup

%% file: files/02introduction.tex
\section{Introduction}

Industry 5.0 is pioneering a new engineering approach, focusing on human-centric solutions and emphasizing the importance of human beings in the systems \citep{indus5.0_XU2021530}. The cognitive and executive functions of operators  are significantly impaired under intense and prolonged hot-humid exposure, potentially leading to more severe secondary disasters \citep{threemile2004human}. Within this context, it is vital to have a comprehensive understanding of the cognitive and reliability process of operators, both psychologically and physiologically, under intense heat stress in extremely hot and humid environments. \par

When individuals are exposed to an environment outside the range of thermal comfort, several typical cognitive and executive functions are significantly affected \citep{65CHEN2020107372}. The vigilant attention, automatic responses, and verbal memory are greatly impaired at 37.8℃ compared to 21.℃, reflected by decreased performance in the serial reaction time test and the accuracy of the Stroop task \citep{62YEOMAN2022103743}. Improved cognitive abilities (including semantic interference and visual perception ability, spatial orientation ability, mental arithmetic ability, perception, attention, concentration, learning speed, short-term and long-term working memory, and arousal level) were observed at lower humidity (50\% RH compared with 70\% RH) \citep{7tian2021decreased}. Furthermore, the participants’ thinking, inattention, fatigue, and emotional levels showed significant improvements under lower humidity, as indicated by the subjective scales \citep{7tian2021decreased}. The accuracy of cognitive functions (i.e., semantic interference, visual perception, spatial positioning, thinking, and arousal level) decreases notably at 37℃ after a 45-minute moderate-intensity exercise, while the speed increased \citep{65CHEN2020107372}. As for risk tolerance, an important element in emergency handling, is affected at 50℃ and 20\% relative humidity, with less risk perception on the same risky behaviors and more risk-taking behaviors \citep{61riskCHANG2017150}. For complex tasks such as perceptual-motor tasks including alertness, operating vehicle equipment, tracking, etc., the operational performance began to decrease significantly in the range of 30°C-33°C wet bulb globe temperature (WBGT) \citep{3ramsey1995}. \par

Several hypotheses and models have been developed to gain a comprehensive understanding of the relationship between heat stress and operators' cognitive performance \citep{4hancock2003}. The Yerkes-Dodson law, which explains the inverse U-shaped relationship between human performance and arousal level, has been migrated to explain the relationship between human performance and the thermal load, as heat stress is a critical factor in arousal level \citep{inverU_provins_environmental_1966}. On this basis, the maximal adaptability model \citep{maximodel_hancock_dynamic_1989} has been proposed to improve validity and robustness. It reveals that a too-cold environment will lead to a lower arousal level and worse performance. When the temperature rises, individuals become more alert and perform better. When the temperature is within a certain suitable range, there is no obvious thermal impact on cognitive performance. This is because psychological and physiological states are within the adaptable range allowing cognitive resources to be dynamically adjusted to resist the influence of the thermal environment without compromising the main task performance. When the temperature keeps increasing, attention resources will be exhausted.  Consequently, the human body is unable to allocate cognitive resources to primary tasks through self-regulation, resulting in a decline in performance. This hypothesis has been confirmed by several experiments. An inverted U-shaped relationship was fitted between the comprehensive heat stress index (reflecting the combination influence of physiological and subjective responses) and the accuracy of cognitive tests, while the speed of cognitive tests showed the opposite trends ($R^2>0.8$) \citep{67expU_LIU2022108431}. Heart rate, a potential biomarker of heat stress, has an inverted U-shaped relationship with the relative accuracy of various cognitive tests \citep{heartU_CHEN2023109801}. A 4th order-function paradigm was fitted between the mean skin temperature and the relative cognitive performance (RCP), with a stable RCP in the range from 36.0℃ to 37.25℃ and deteriorated RCP out of this range \citep{60expU_ZHU2023112704}.
\par

With the advancements in non-invasive neurophysiological measurement, particularly in electrocardiogram (ECG) \citep{zhu2023cognitive}, functional near-infrared spectroscopy (fNIRS) \citep{39pan2019applications}, eye tracking \citep{32tran2017predicting}, electroencephalogram (EEG) \citep{kim2020psychophysiological}, electromyography (EMG) \citep{1malmo2000electromyographic}, and electrodermal activity (EDA) \citep{52visnovcova2016complexity}, it is now possible to investigate the process and mechanism of performance impairment under severe heat stress with greater insight. Higher heat rate \citep{64ABBASI2020103189, 65CHEN2020107372}, a higher ratio of low frequency to high frequency (LF/HF) of ECG signals \citep{63LAN201029, 64ABBASI2020103189}, and a decrease in pNN50 (percentage of successive RR intervals that differ by more than 50 milliseconds) \citep{65CHEN2020107372} were noted with increased exposure to the hot-humid environment, associated with more intense activity in the autonomic nervous system and worse regulation capability of cardiovascular system. Lower peak amplitudes of the P300 component \citep{66nakata_effects_2021} and lower theta-band EEG \citep{63LAN201029} revealed the impairment of neural activity and lower motivation in cognitive function under heat stress. The pupil diameter is positively correlated with fatigue level and was minimized at 24℃ and 40\% relative humidity, while the minimized standard deviation of pupil diameter indicated an optimal concentration level \citep{68pupil_LIU2021101458}. Several brain regions were suppressed by passive hyperthermia, including the bilateral motor cortex and left lateral-occipital cortex (compromised visual processing) \citep{fmri1_TAN2023}. However,  enhanced activation was reported in some other regions, such as the right superior frontal gyrus, temporal lobe, and right intra-parietal sulcus \citep{fmri3_LIU2013220, fmri4_jiang_hyperthermia_2013}. The hyperthermia effect on the dorsolateral prefrontal cortex (DLPFC), which is a vital brain region related to executive functions, working memory, cognitive control, and emotion regulation, remains controversial. Some studies reported enhanced activity \citep{fmri3_LIU2013220, fmri4_jiang_hyperthermia_2013}, while others reported decreased activity \citep{fmri2_qian_disrupted_2020}. Furthermore, the reduced functional connectivity within the default mode network provides another insight into the impairment of executive control performance \citep{fmri2_qian_disrupted_2020}.  \par

Previous studies have extensively explored the qualitative relationship between heat strain and performance, as well as the potential biological indicators that varied with heat stress and their possible biological explanation. However, there are still research gaps. Most research has focused on temperatures below 35℃ WBGT or 38℃ and lacks evidence under extreme conditions encountered in serious disasters. Furthermore, there is still a lack of research on mechanisms supported by multi-physiological measurement methods.\par

Therefore, this work aims to investigate the effect of extreme heat stress on operators' performance and typical cognitive abilities required to deal with emergencies, such as situation awareness and work memory. Moreover, flexible physiological measurement methods with high reliability and validity including ECG, fNIRS, and eye tracking were employed to understand the mechanism underlying performance impairment. The rest of the paper is organized as follows. Section 2 illustrates the experiment, including major tasks, psychological measurements (situation awareness, working memory, and workload), and physiological measurements (fNIRS, ECG, and eye tracking), along with their data analysis methods applying neuroscience domain knowledge. In section 3, t-tests, analysis of variance, ECG signal time spectrums, task-related fNIRS signal activation analysis, rest-stating fNIRS signal functional connectivity analysis, and eye-tracking data heat map analysis were conducted to investigate the impact of hot-humid exposure on cognitive functions and human performance. In section 4, the physiological and psychological mechanisms of impaired cognitive functions under intense heat stress were discussed. The neural activity (ECG), energy metabolism (fNIRS), and visual search patterns (eye tracking) were explored, providing a neurophysiological basis for the modified maximal adaptability model. In section 5, the limitations and contributions are discussed.

%% file: files/03method.tex
\section{Methodology}

\subsection{Approach}

\begin{figure}[H]
\centering
\subfloat[]{\includegraphics[width=0.7\linewidth]{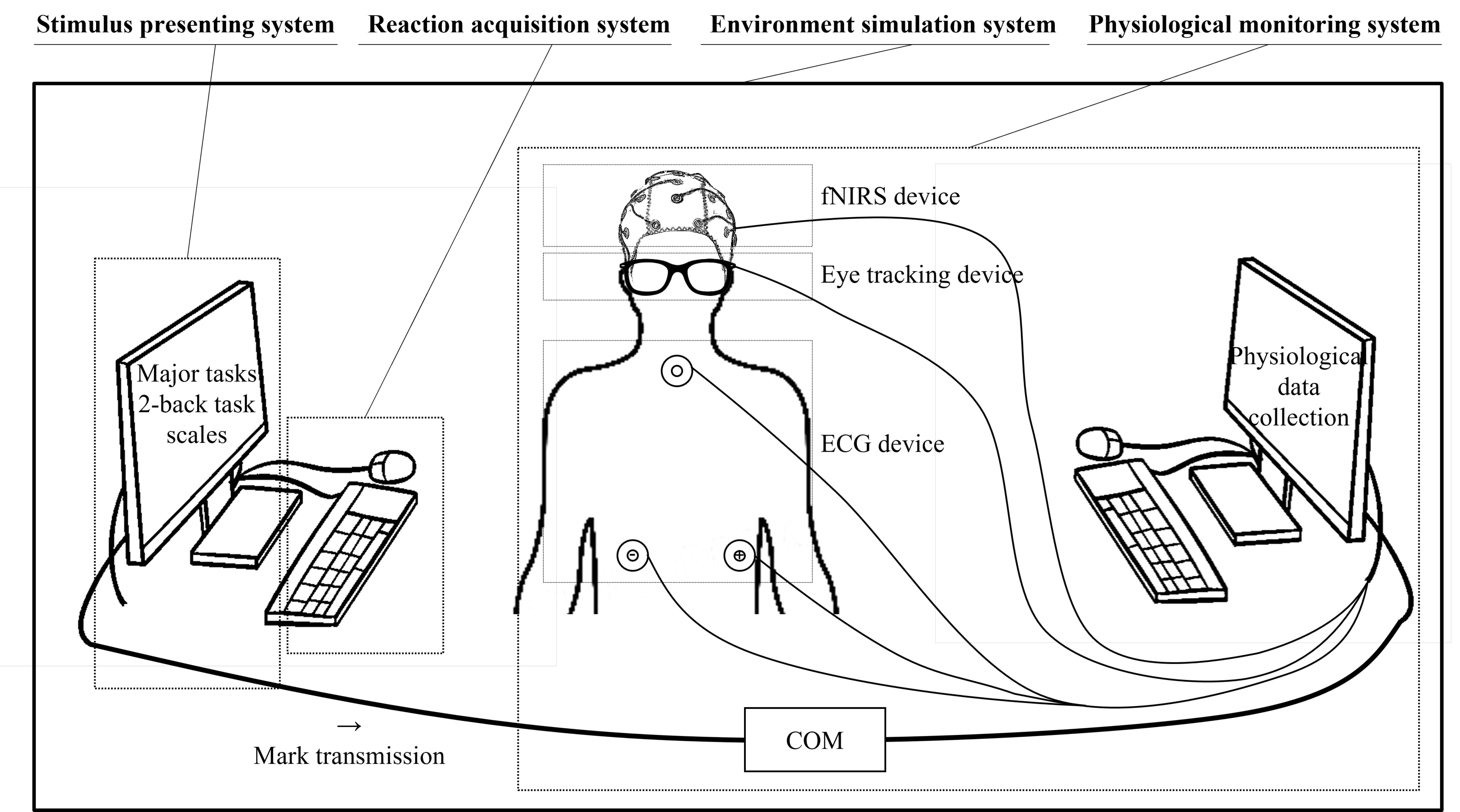}}\\
\subfloat[]{\includegraphics[width=0.7\linewidth]{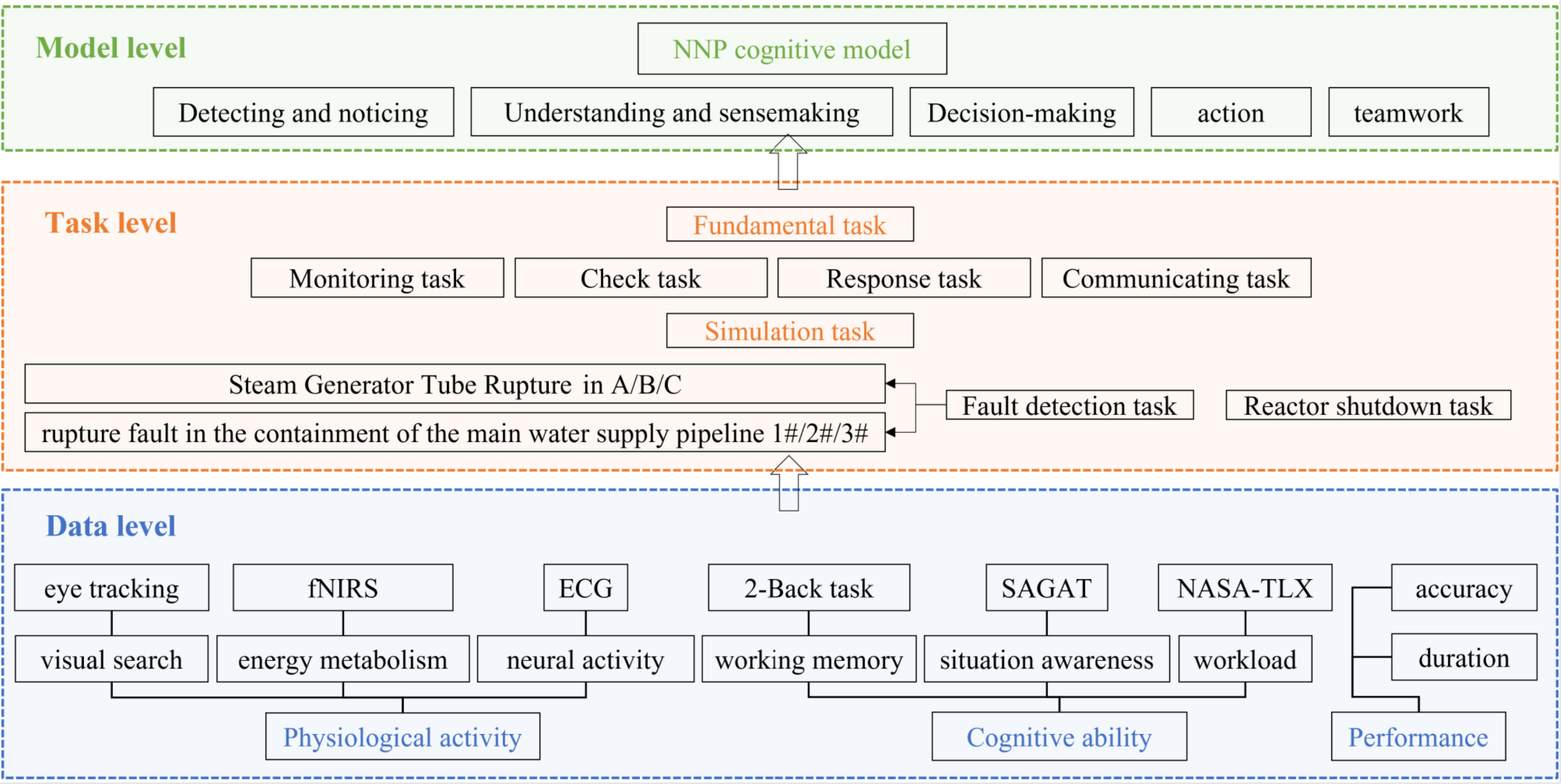}}\\
\caption{Illustration of experimental design. (a) schematic diagram of the experiment setup. The environment simulation system regulates temperature and humidity to achieve the desired WBGT. The stimulus presentation system displays NPPs fault detection tasks, response tasks, and questionnaires that participants interact with. The reaction acquisition system collects performance data including mouse and keyboard input. The physiological monitoring system contains an fNIRS device (on the forehead), an eye-tracking device (glasses), an ECG device (on the chest), and a computer that collects physiological data and markers from the stimulus computer via COM. (b) outline of the experiment. The experimental method includes three levels from macroscopic to concrete: model level, task level, and data level. The model level summarizes cognitive processes and typical abilities operators need. The task level sets fundamental tasks that cover elements mentioned in the model level and selects typical NPPs tasks corresponding to those fundamental tasks. The data level seeks specific methods and indicators from tasks to quantify performance, cognitive ability, and physiological activity reflecting operators’ mental state.}
\label{fig:exp setting}
\end{figure}

In the climate lab, thirty participants were exposed for a total of 120 minutes (30 minutes $\times$ 4) under four different conditions: 25℃ 60\% RH (21.26 ℃ WBGT), 30℃ 70\% RH (26.88 ℃ WBGT), 35℃ 80\% RH (31.92 ℃ WBGT), and 40℃ 90\% RH (36.81 ℃ WBGT). The participants' performance was recorded during major tasks, including typical NPPs  (nuclear power plants) monitoring, checking, and operating tasks extracted from classic NPPs operators’ cognitive model. Cognitive abilities were measured by response tasks objectively and by self-rating questionnaires subjectively. Physiological activities were recorded in real-time using an fNIRS device, eye-tracking glasses, and an ECG device placed on the participant's forehead, eyes, and chest respectively. The experimental design is shown in \Cref{fig:exp setting}.

\subsection{Subjects}

Thirty volunteers (16 females and 14 males) participated in this experiment, aged 20-39 years (mean: 26±5 years). Prior to the experiment, all subjects were well-trained and tested to minimize the effects of practice. They were instructed to refrain from consuming caffeine and to get sufficient sleep the night before. Informed consent was obtained from all participants.

\subsection{Experimental environment}
To simulate a high temperature and humidity environment caused by emergencies, the experiment was implemented in the constant temperature and humidity laboratory, which is capable of adjusting the temperature within the range of 20℃ to 80℃ (±0.8℃) and the humidity within the range of 30\% to 95\% (±5\%). Additionally, it provides a working space measuring 6m*5m*4.7m. \par
% supported by the Hefei Institute for Public Safety Research, Tsinghua University
In this experiment, we intended to explore the coupling effect of high temperature and high humidity on NPPs operators’ performance. Therefore we selected the wet bulb globe temperature index (WBGT) as the independent variable. WBGT is an estimation of the heat stress on workers \citep{0iso}, which has been widely used as a guideline to determine appropriate work intensity for operators in hot and humid conditions. The experiment consists of four conditions and the parameters of each condition are shown in \Cref{table:condition_setting}. %Indoor, WBGT is calculated by the following formula, in which $t_{nw}$ is the natural wet bulb temperature, and $t_g$ is the black globe temperature.

%\begin{equation}
%    \textrm{WBGT}=0.7t_{nw}+0.3t_g
%\end{equation}

%Moreover, WBGT can be estimated by the dry-bulb temperature (Ta), and the relative humidity (RH) according to the following formula:

%\begin{equation}
%    \textrm{WBGT}=0.707*\textrm{Ta}+0.09*\textrm{RH}.
%\end{equation}

%\begin{table}[H]
%    \centering
%    \caption{Condition parameters}
%    \begin{tabular}{ccccc}
%        \toprule
%        Condition & Ta(℃) & RH(\%) & WBGT theoretical(℃) & WBGT measuring (℃) \\ 
%        \midrule
%        1 & 25 & 60 & 23.075 & 21.26±0.40  \\ 
%        2 & 30 & 70 & 27.51 & 26.88±0.85  \\ 
%        3 & 35 & 80 & 31.945 & 31.92±0.41  \\ 
%        4 & 40 & 90 & 36.38 & 36.81±0.89  \\ 
%        \bottomrule
%    \end{tabular}
%    \label{table:condition_setting}
%\end{table}

\begin{table}[H]
    \centering
    \caption{Condition parameters}
    \begin{tabular}{cccc}
        \toprule
        Condition & Ta(℃) & RH(\%) & WBGT (℃) \\ 
        \midrule
        1 & 25 & 60 & 21.26±0.40  \\ 
        2 & 30 & 70 & 26.88±0.85  \\ 
        3 & 35 & 80 & 31.92±0.41  \\ 
        4 & 40 & 90 & 36.81±0.89  \\ 
        \bottomrule
    \end{tabular}
    \label{table:condition_setting}
\end{table}

\subsection{Major task}\label{sec:major task}

According to NPPs operators’ cognitive model \citep{cogmodel1ohara2008human, cogmodel2whaley2016cognitive}, detecting and noticing, understanding and sensemaking, decision-making, action, and teamwork are the five essential cognitive and executive functions that operators require. To evaluate these essential functions, we designed four fundamental tasks: monitoring, checking, responding, and communicating respectively. \par
In the specific NPPs scenario, two kinds of fault detection tasks were selected: SGTR (Steam Generator Tube Rupture) (fault1) and rupture fault in the containment of the main water supply pipeline (fault2). These tasks encompass monitoring, checking, and communicating. The executive task of reactor shutdown was selected to cover checking, responding, and communicating.\par
To simulate NPPs operators’ tasks under emergencies, we developed a virtual central control room software based on the MATALB app designer. Fault detection tasks and reactor shutdown task were implemented in the software. Subjects were asked to complete three major tasks according to the emergency plan as quickly and accurately as possible in the software. Meanwhile, duration, accuracy, error counts, fault detection time, and fault confirmation time were recorded automatically to assess the operators’ performance. \par

\begin{itemize}
    \item Major task 1: detect fault1 and confirm the fault type and location.
    \item Major task 2: detect fault2 and confirm the fault type and location.
    \item Major task 3: confirm the key parameters and shut down the reactor.
\end{itemize}

\subsection{Psychological measurement}

To investigate operators’ mental state and cognitive ability in the task under emergencies, situation awareness, working memory, and workload were measured. 

\subsubsection{Situation awareness (SA)}
%\paragraph{Situation awareness (SA)}
SA reflects operators’ perception of vital parameters, the state of key controllers, and their comprehension of the NPPs operating state at any given time. It also includes the projection of the vital systems’ status in the near future. An operator with good SA can confirm faults and respond to emergencies quickly, thereby avoiding additional losses.
SAGAT \citep{SAGATendsley1988situation} (Situation Awareness Global Assessment Technique) has been widely used to measure SA which has demonstrated good reliability and validity. SAGAT questions in this experiment were selected as:

\begin{itemize}
    \item Perception
    \begin{itemize}
        \item Write the approximate range of vital parameters (pressure, temperature, flow, liquid/material level, etc.)
        \item Write the state of the key controller (lamp, switch, valve, pump, etc.)
    \end{itemize}
    \item Comprehension
    \begin{itemize}
        \item Judge the status of key equipment (reactor, primary circuit, secondary circuit, steam generator, steam turbine, etc.)
        \item Judge the status of key systems (coolant injection system, auxiliary water supply system, etc.)
        \item Judge the occurring accident and locate the fault.
    \end{itemize}
    \item Projection
    \begin{itemize}
        \item Predict whether key equipment and systems can work after the accident.
        \item Predict how would the status change after intervention.
    \end{itemize}
\end{itemize}

\subsubsection{Working memory}

\begin{figure}[H]
    \centering
    \includegraphics[width=0.35\linewidth]{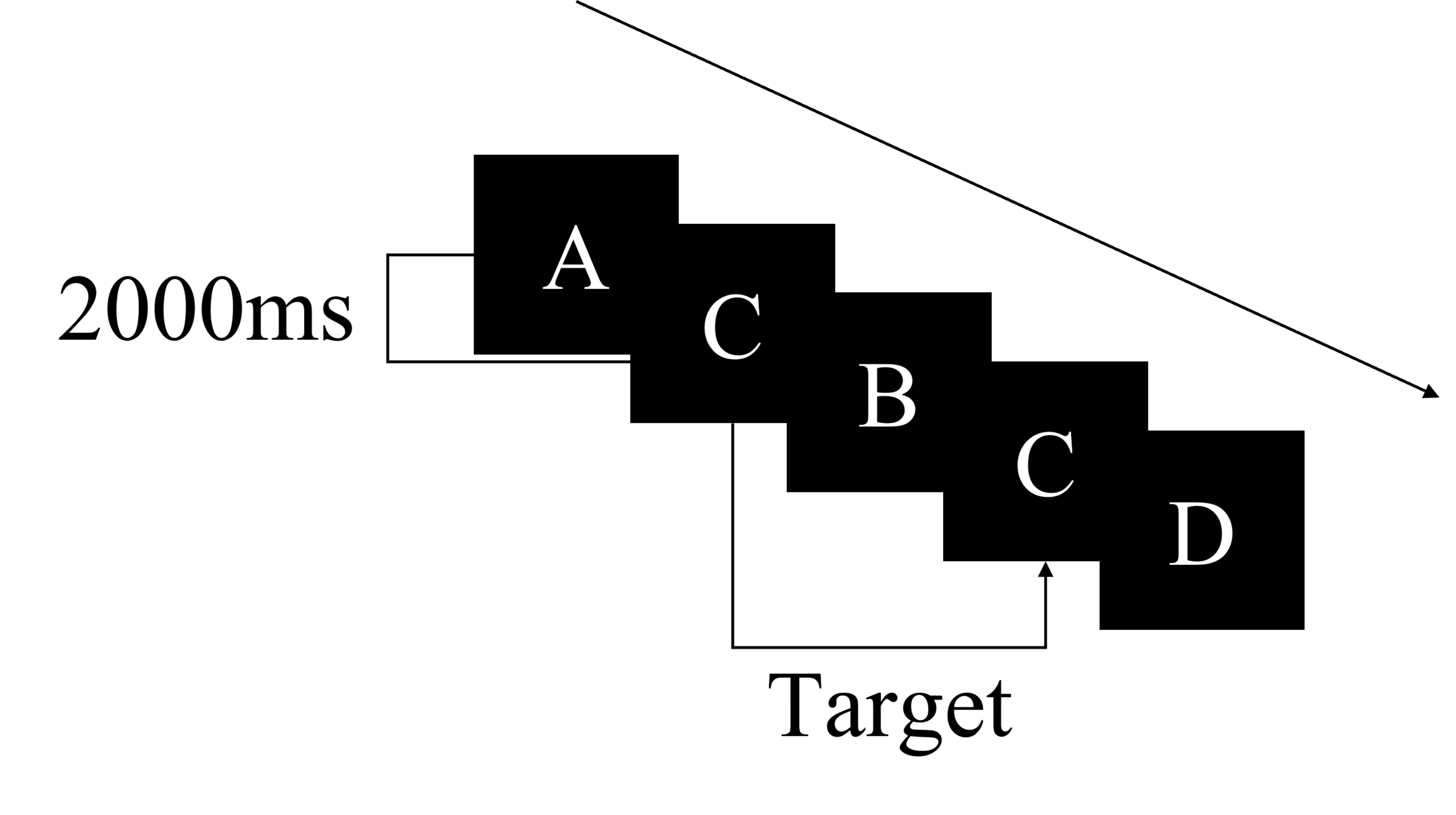}
    \caption{The experiment flow for 2-back working memory task}
    \label{fig:2back_setting}
\end{figure}

When emergencies occur, NPPs operators must process an enormous information flow. They need to extract crucial details and temporarily store them in working memory, which is an important brain executive function. Working memory was measured by a 2-back task developed by psychtoolbox3 in MATLAB. A continuous sequence of letters was displayed on the screen, and the subjects should press the arrows on the keyboard based on the comparison of each letter with the previous second one (same-left, different-right) (\Cref{fig:2back_setting}). The task lasted for 3 minutes and included 30 trials. 

\subsubsection{Workload}
The workload can be divided into two aspects: physiological load and mental load. There is an inverse U relationship between performance and workload. Operators might become bored and not alert in emergencies when the workload remains at a low level for a long time. However, they might also find it difficult to concentrate on the major task when their remaining cognitive resource has been consumed quickly due to prolonged workload. The NASA task load index (NASA-TLX) scale was applied to assess operators’ workload subjectively including mental demand, physical demand, temporal demand, performance, effort, and frustration.

\subsection{Physiological measurement}
\subsubsection{ECG}
ECG signal was recorded by a 3-lead ECG module (BIOPAC MP160, BIOPAC Systems Inc., Santa Barbara, CA), which provides information about the regulation of the cardiovascular system by neurohumoral factors and reflects the balance of activity intensity between the sympathetic and vagal nerves.
\subsubsection{fNIRS}
\begin{figure}[H]
    \centering
    \includegraphics[width=0.5\linewidth]{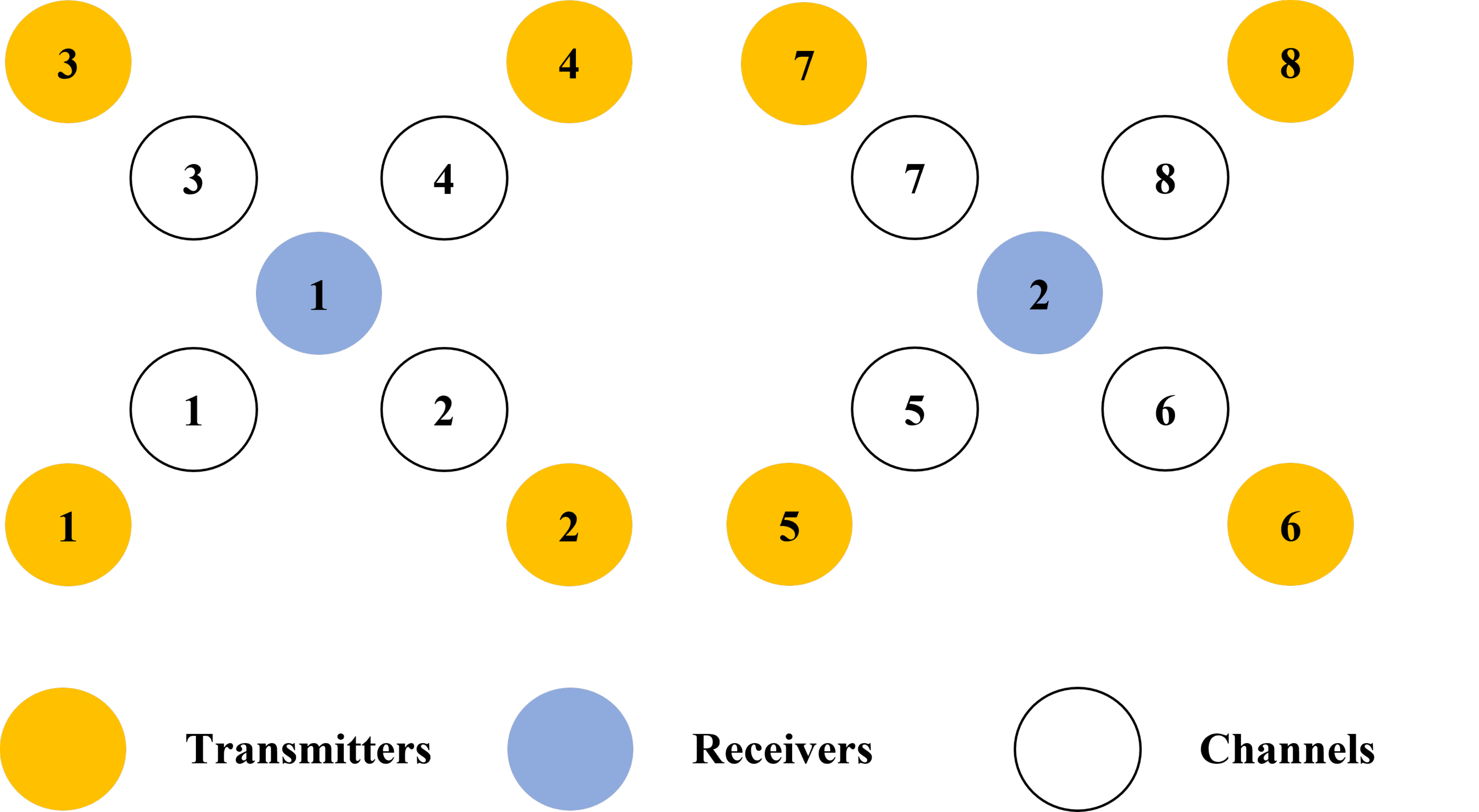}
    \caption{fNIRS probe placement}
    \label{fig:fnirs_setting}
\end{figure}

fNIRS is a non-invasive neuroimaging method that measures blood oxygenation, that reflects the energy metabolism of specific brain areas. High hemoglobin concentration (HBO) is typically associated with activation of brain regions. The prefrontal cortex (PFC) is an important brain region associated with various executive functions that we are interested in, including cognition, decision-making, and working memory. An 8-channel fNIRS device (Artinis Octamon, Netherlands) was placed on the PFC to measure brain activity, with the probe placement shown in \Cref{fig:fnirs_setting} and \Cref{table:fnirs_ba}. In addition, the NIRS-KIT toolbox \citep{nirskit} was utilized to assist in processing the fNIRS data partly. 

\begin{table}[H]
    \centering
    \caption{Localization of fNIRS  channels and their corresponding  Brodmann areas.}
    \begin{tabular}{ll}
        \toprule
        channel  & Brodmann area  \\ 
        \midrule
        ch1   &  10 - Frontopolar area \\
            & 46 - Dorsolateral prefrontal cortex \\ 
        ch2  & 10 - Frontopolar area   \\ 
        ch3  & 45 - pars triangularis Broca’s area  \\
            & 46 - Dorsolateral prefrontal cortex \\ 
        ch4  & 9-46 - Dorsolateral prefrontal cortex \\ 
            & 10 - Frontopolar area   \\ 
        ch5 &  10 - Frontopolar area \\
            & 46 - Dorsolateral prefrontal cortex  \\ 
        ch6   & 45 - pars triangularis Broca’s area  \\
            & 46 - Dorsolateral prefrontal cortex   \\ 
        ch7  & 9-46 - Dorsolateral prefrontal cortex \\ 
            & 10 - Frontopolar area   \\ 
        ch8  & 45 - pars triangularis Broca’s area  \\
            & 46 - Dorsolateral prefrontal cortex  \\ 
        \bottomrule
    \end{tabular}
    \label{table:fnirs_ba}
\end{table}

\subsubsection{Eye Tracking}
Eye tracking data was obtained by Dikablis Glass 3 to observe the distribution of participants’ attention during tasks. This allowed the exploration of operators’ virtual search patterns in hot and humid environments. Prior to each condition, eye movement calibration was performed. 

\subsection{Procedure}
The study involved four different WBGT conditions, and each consists of nine phases: baseline rest (P1), completion of major task1 and answering SAGAT questions (P2-3), completion of major task2 and answering SAGAT questions (P4-5), completion of major task3 and answering SAGAT questions (P6-7), answering the NASA-TLX scale (P8), completion of the 2-back task (P9), recovery while waiting for temperature and humidity to reach the target, and looping through all the phases again until the tasks in all the conditions were completed. The major tasks were outlined in \S\ref{sec:major task} but their order was randomized.

\begin{figure}[H]
    \centering
    \includegraphics[scale=0.5]{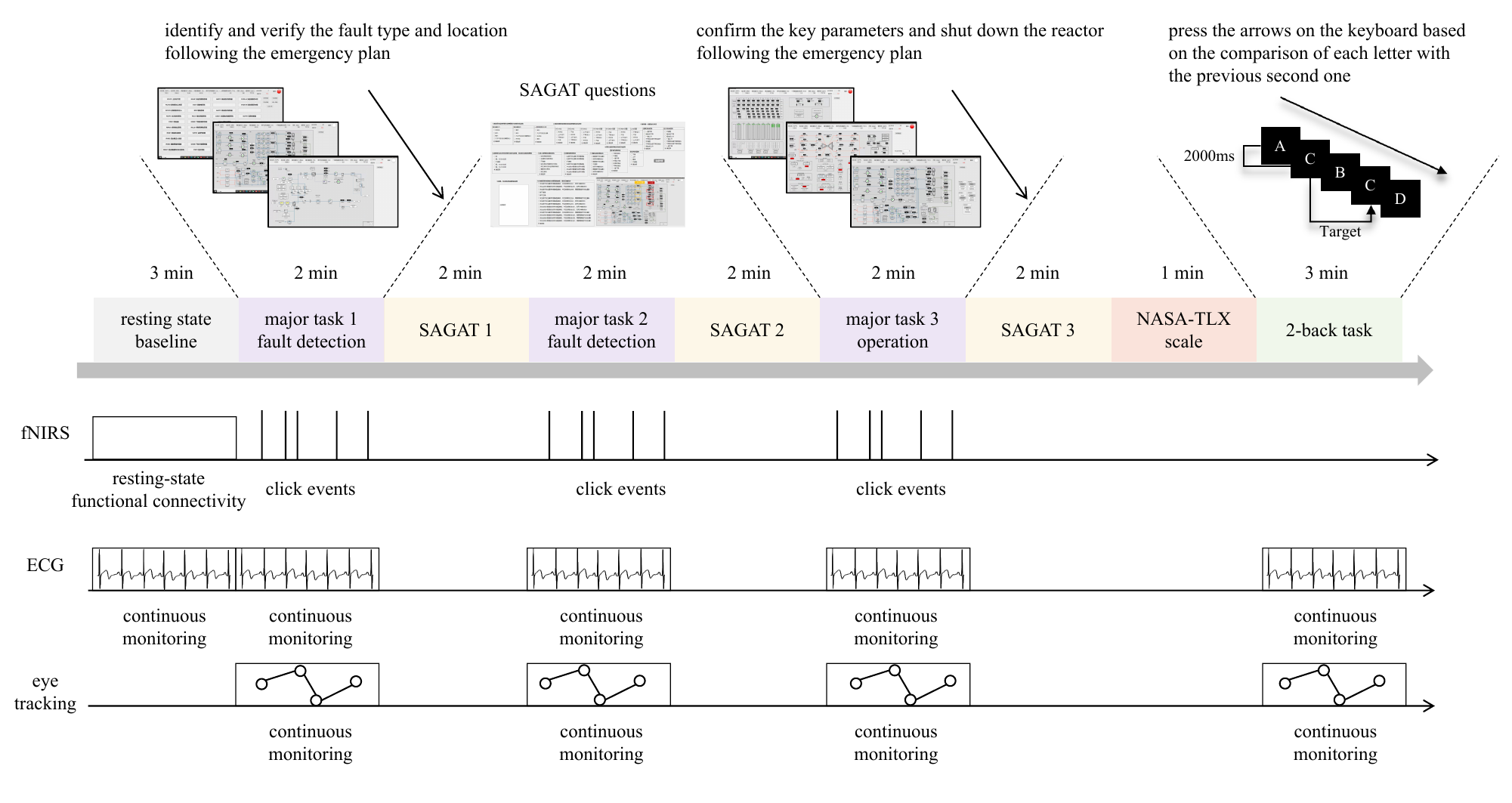}
    \caption{Experiment procedure in every condition. Each phase lasted for 5 minutes.}
    \label{fig:exp_flow}
\end{figure}

\subsection{Data analysis}
\subsubsection{Performance and psychological measurement data}
Performance and psychological data were recorded by the developed software mentioned in \S\ref{sec:major task} automatically, and the extracted features for each task are listed in \Cref{table:task_features}. \par

The difference of each feature for each subject was calculated between adjacent conditions (condition1 and 2 , 2 and 3, 3 and 4). One sample t-test was performed to determine any significant difference from 0.
 % font size
\begin{table}[H]
    \centering
    \caption{Performance and psychological measurement features}
    \begin{tabular}{p{3cm}p{3.2cm}p{2.8cm}p{6.6cm}}
        \toprule
        performance & Situation awareness & workload & Work memory  \\ 
        \midrule
        task completion rate [\%] & SAGAT completion time [s] & mental demand score & number of correct responses [n.u.]  \\ 
        error numbers [n.u.] & total score & physical demand score & number of wrong responses [n.u.]  \\ 
        fault detection time [s] & level1 perception sore & temporal demand score & number of missed responses [n.u.]  \\ 
        fault confirmation time [s] & level2 comprehension score & performance score & mean reaction time of all responses [s]  \\ 
        completion time [s] & level3 projection score & effort score & standard deviation of all response reaction times [s]  \\ 
        ~ & communication score & frustration score & mean reaction time of all correct response [s]  \\ 
        ~ & ~ & total score & standard deviation of all correct response reaction times [s]  \\ 
        \bottomrule
    \end{tabular}
    \label{table:task_features}
\end{table}

\subsubsection{ECG and HRV}

\textit{\textbf{Time domain analysis}}. Signal filtering and the QRS complex detection were applied to extract the RR interval series. From these series, various characteristic features were calculated and summarized in \Cref{table:ecg feature}. Analysis of variance (ANOVA) was performed on each feature to identify significant differences between any two conditions. \par
\textit{\textbf{Frequency domain analysis}}. The fast Fourier transform (FFT) was applied to the filtered ECG data to calculate its power spectral density (PSD). Spectral power was calculated as the area below the PSD curve which was then divided into four frequency bands: very low-frequency power (VLF) in [0, 0.04] Hz, low-frequency power ( LF) in [0.04, 0.15] Hz, high-frequency power (HF) in [0.15, 0.4] Hz and very high-frequency power (VHF) in [0.4, 3] Hz. Typical features are summarized in \Cref{table:ecg feature}. ANOVA was also performed on the frequency features. \par

\begin{table}[H]
    \centering
    \caption{ECG features in time domain and frequency domain.}
    \begin{tabular}{p{1.5cm} p{3.4cm} p{1.2cm} p{9cm}}
        \toprule
        Domain  & ECG features & Unit  & Description   \\ 
        \midrule
        \multirow[t]{13}{1.5cm}{Time domain} & RR interval & [s] & Average of time distance between two adjacent R peaks  \\ 
        ~ & HR & [bpm] & Numbers of heart beat every minute   \\ 
        ~ & R amplitude & No unit & R wave peak amplitude   \\ 
        ~ & P amplitude & No unit & P wave peak amplitude  \\ 
        ~ & QRS width & [s] & Time between onset and end of the QRS complex  \\ 
        ~ & PRQ width & [s] & Time between onset of the P wave to the Q wave  \\ 
        ~ & QT interval & [s] & Time between the beginning of the Q wave and the end of the T wave  \\ 
        ~ & QTC interval & [s] & QT time interval divided by the square root of the RR interval  \\ 
        ~ & ST interval & [s] & Time between the S wave to the end of the T wave  \\ 
        ~ & SDNN & [s] & Standard deviation of RR time series    \\ 
        ~ & RMSSD & [s] & Root mean square of the difference of all subsequent RR intervals  \\ 
        ~ & SDSD & [s] & Standard deviation of the difference of all subsequent RR intervals  \\ 
        ~ & PNN50 & \% & Percentage of RR intervals in which the change of successive NN exceeds 50 ms  \\ 
        \multirow[t]{7}{1.5cm}{Frequency domain} & VLF  & [$\mathrm{s}^2$] & Spectral power of the RR time series in the band [0; 0.04 Hz]  \\ 
        ~ & LF  & [$\mathrm{s}^2$] & Spectral power of the RR time series in the band [0.04; 0.15 Hz]  \\ 
        ~ & HF  & [$\mathrm{s}^2$] & Spectral power of the RR time series in the band [0.15; 0.4 Hz]  \\ 
        ~ & VHF  & [$\mathrm{s}^2$] & Spectral power of the RR time series in the band [0.4; 3 Hz]  \\ 
        ~ & Sympathetic-vagal balance & No unit & Ratio between LF and HF  \\ 
        ~ & Sympathetic & No unit & Ratio between LF and (VLF+LF+HF)  \\ 
        ~ & Vagal & No unit & Ratio between HF and (VLF+LF+HF) \\ 
        \bottomrule
    \end{tabular}
    \label{table:ecg feature}
\end{table}

\textit{\textbf{Time-frequency domain analysis}}. The ECG data was a non-stationary signal in every task and condition. Therefore, all ECG data was scaled to [0, 100\%] on the timeline to obtain the same-length ECG data and the Short-time Fourier Transform (STFT) with [0, 3Hz] frequency limits, 0.5 frequency resolution, 20\% overlapping, 0.875 leakage function was applied on the filtered ECG to get time spectrogram. Time-frequency representation of the RR time series provides valuable insights into the changes in frequency components and energy distribution over time. 

\subsubsection{fNIRS}
Raw fNIRS data contains several types of noise. Thus, the fNIRS signal was bandpass filtered between 0.01Hz and 0.1Hz to eliminate 50Hz noise as well as most physiological noise. First detrending was implemented to prevent baseline drift. Motion correction was then applied to mitigate motion artifacts. Following preprocessing, a general linear model analysis was conducted to plot the PFC activation map and identify valid activated brain regions under tasks. Additionally, functional connectivity analysis was conducted on the resting-state fNIRS data to explore brain synchrony in different conditions. \par
\textit{\textbf{General linear model (GLM) activation analysis}}. Beta values were derived by regressing the HBO curve against the reference curve, which was obtained by convolving the click event list with the HRF function. A one-sample t-test (different from 0) was performed on the betas to figure out the activated channels in the PFC. \par
\textit{\textbf{Resting-state functional connectivity (FC) analysis}}. The correlation coefficient (CORR), correlation coefficient after fishier-Z transformation (CORR fisher-Z), coherence (COH), and phase-locking value (PLV) were calculated between different channels. The results were corrected by multiple comparisons using NBS and a functional connectivity matrix was generated. \par
The correlation coefficient between the signal x and y is calculated by the following formula:
\begin{equation}
    R_{xy}=\frac{1}{N}\sum_{k=1}^{N}x(k)y(k)
\end{equation}

Coherence is calculated by the following formula, in which $S_{xy}(f)$ is cross-power spectral density between signal $x$ and $y$ at frequency $f$, $S_{xx}(f)$ and $S_{yy}(f)$ are auto-power spectral density for signal $x$ and $y$ at frequency $f$ respectively.
\begin{equation}
    \mathrm{COH}_{xy}=|K_{xy}(f)|^2=\frac{|S_{xy}(f)|^2}{S_{xx}(f)S_{yy}(f)}
\end{equation}

Phase-locking value is calculated by the following formula, in which $\Delta\phi_{rel}(t)$ is the phase difference between the complex signal $\Tilde{x}$ and $\Tilde{y}$ obtained by the Hilbert transform.
\begin{equation}
    \mathrm{PLV}=|\langle e^{i\Delta\phi_{rel}(t)}\rangle|=\bigg|\frac{1}{N}\sum_{n=1}^N e^{i\Delta\phi_{rel}(t_n)}\bigg|=\sqrt{\langle\cos\Delta\phi_{rel}(t)\rangle^2+\langle\sin\Delta\phi_{rel}(t)\rangle^2}
\end{equation}

\subsubsection{Eye tracking}
\begin{figure}[H]
    \centering
    \includegraphics[width=1\linewidth]{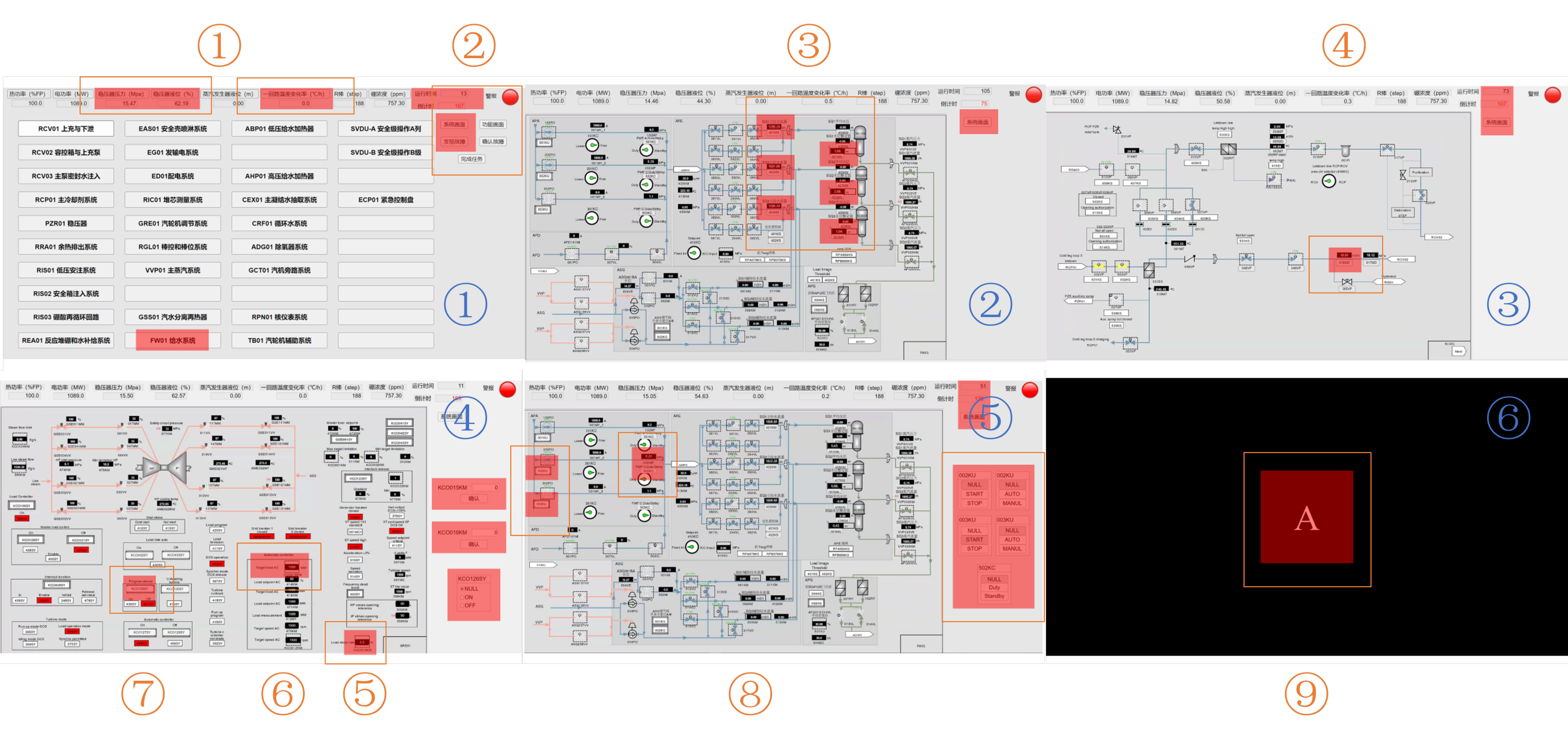}
    \caption{AOIs settings. Red areas are AOIs that cover important information that operators should monitor, check, or control.}
    \label{fig:aoi setting}
\end{figure}

\textit{\textbf{AOI and pupil analysis}}. 6 pages of interest and 9 areas of interest (AOIs) were set up (shown in \Cref{fig:aoi setting}) to collect eye-tracking related features (listed in \Cref{table:eye tracking feature}) which can reflect information search efficiency, information processing efficiency, mental load, and concentration.
The difference of each feature in each subject was calculated between the adjacent conditions (condition1 and 2 , 2 and 3, 3 and 4). One sample t-test was performed to identify a significant difference from 0.

\begin{table}[H]
    \centering
    \caption{Eye-tracking related features}
    \begin{tabular}{p{2cm} p{4cm} p{1.2cm} p{8cm}}
        \toprule
        Domain & Feature  & Unit  & Description   \\ 
        \midrule
        \multirow[t]{9}{2cm}{AOI related}  & Number of glances$>$2s & No unit & The number of single glance more than 2 seconds  \\ 
        ~ & Total Glance time  & [s] & Total glance time on a specific AOI  \\ 
        ~ & Mean Glance Duration  & [s] & Average duration of each glance, total glance time / number of glances\\ 
        ~ & Glance Rate  & [1/s] & The number of glances on a AOI every second  \\ 
        ~ & AOI Attention Ratio & [\%] & Glance time on a specific AOI / task duration  \\ 
        ~ & Vertical Eye Activity  & [pixel] & Standard deviation of y coordinates in the screen   \\ 
        ~ & Mean fixation duration left  & [ms] & Average fixation duration of left eye in an AOI  \\ 
        ~ & Mean fixation duration right  & [ms] & Average fixation duration of right eye in an AOI  \\ 
        ~ & Number of saccades right & No unit & Total number of saccades of right eye in an AOI  \\ 
        \multirow[t]{4}{2cm}{Task and pupil related} & Duration & [s] & Duration on an interest page  \\ 
        ~ & Percentage Transition Times & [\%] & Time of transiting between different AOIs / duration  \\ 
        ~ & Pupil left avg & ~ & Average pupil diameter of left eye  \\ 
        ~ & Pupil right avg & ~ & Average pupil diameter of right eye \\  
        \bottomrule
    \end{tabular}
    \label{table:eye tracking feature}
\end{table}

\textit{\textbf{Heat map analysis}}. The heat map was created on 6 pages of interest, displaying a deeper color (red) in areas with more glances and a lighter color (green) in areas with fewer glances. This visualization effectively reflects the distribution of attention.

%% file: files/04result.tex
\section{Results}

\subsection{Task performance and cognitive ability}
\begin{figure}[H]
    \centering
    \includegraphics[width=1\linewidth]{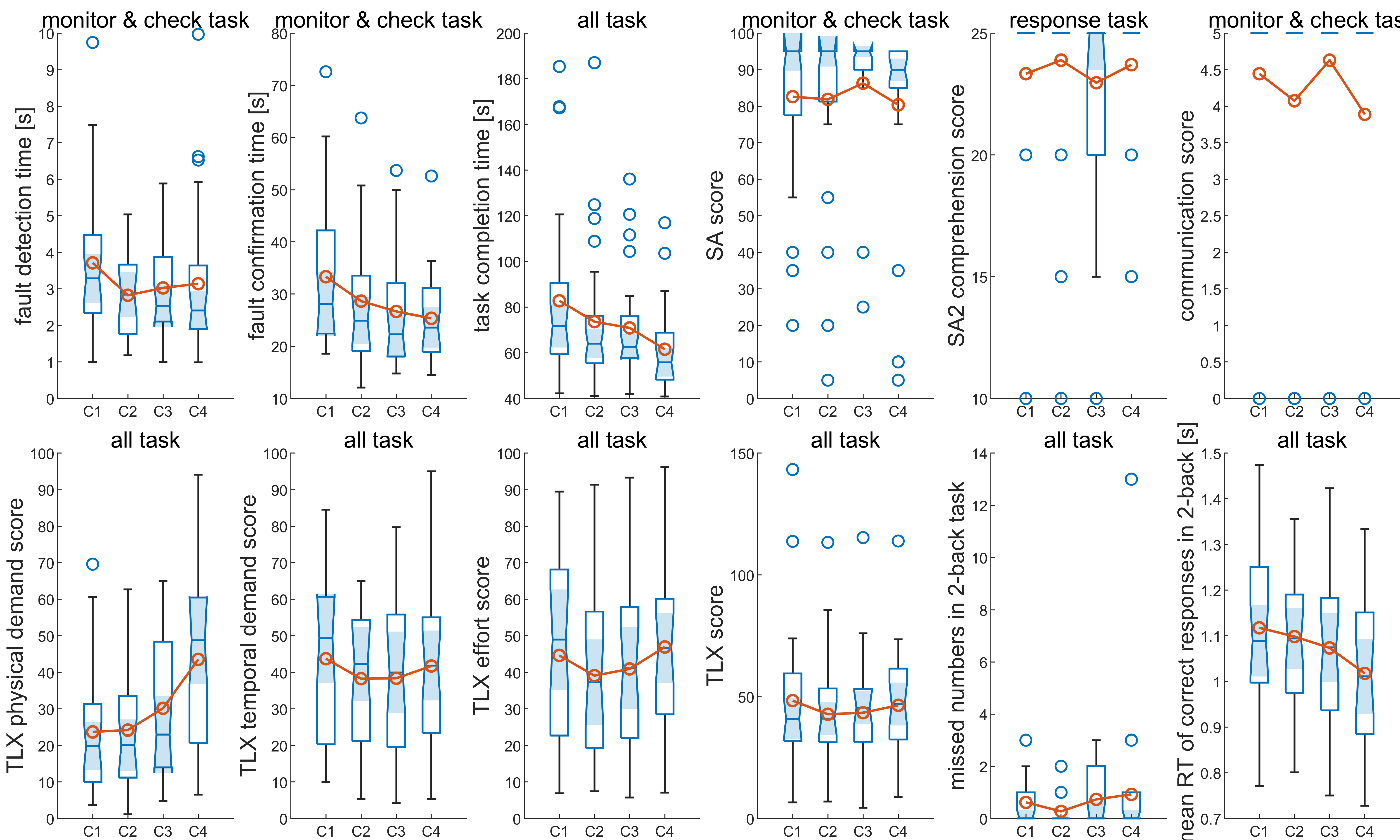}
    \caption{The boxplots display significant results in performance, situation awareness, workload, and work memory. Red lines represent average values.}
    \label{fig:task result}
\end{figure}

One sample t-test (differ from 0) was applied among all the features across adjacent conditions. All the significant p values are shown in \Cref{table:task result} and visualized as boxplots with average lines in \Cref{fig:task result}. Statistical significance is accepted at $p<0.05$.

\subsubsection{Task performance}
Statistic results indicated that for task complement rate, no significant difference was observed between different conditions. Operators can follow all the necessary steps even in a hot and humid environment. \par
In the monitor and check task, the number of misoperations showed no significant result, however in the response task did (between condition 2 and 3, $t=2.59, p=0.02$). It suggested that monitor and check ability was less affected by temperature and humidity. Workers were more likely to make mistakes at around 32℃ WBGT (condition 3). \par
In the monitor and check task, fault detection time in condition 2 was significantly less than condition 1 ($t=-2.11, p=0.05$), as well as fault confirmation time ($t=-3.19, p<0.01$). \par
In response task, task completion time in condition 4 was significantly less than condition 3 ($t=-3.14, p<0.01$), and in condition 2 is significantly less than condition 1 ($t=-2.08, p=0.05$).

\subsubsection{Situation awareness}
As for the total SA score, the score in condition 4 was significantly lower than condition 3 ($t=-2.31, p=0.03$).  \par
For  more detail, in terms of level 1 SA perception, there was no significant difference between different conditions and tasks. It indicated that perception is a low-level situation awareness ability that was less affected by the environment. As for level 2 SA comprehension, there was a significantly lower score in condition 3 than 2 in the response task ($t=-2.43, p=0.02$). As for level 3 SA projection, there was also no significant difference between conditions and tasks. Regarding communication, the score in condition 4 was significantly less than condition 3 ($t=-2.13, p=0.04$). 

\subsubsection{Workload}
In the NASA-TLX scale, physical demand had a significant increase from condition 2 to condition 4 ($t=2.66, p=0.01$ and $t=5.99, p<0.01$). Temporal demand in condition 2 was greatly less than condition 1 ($t=-2.27, p=0.03$). Effort score in condition 4 was greatly higher than condition 3 ($t=2.25, p=0.03$). The task load index total score in condition 2 was greatly less than condition 1 ($t=-2.27, p=0.03$). Moreover, no significant difference was observed in mental demand, performance, and frustration between different conditions.

\subsubsection{Work memory}
In the 2-back task, no great difference was shown in the number of correct responses and wrong responses, but the missed number in condition 2 was greatly less than condition 1 and condition 3 ($t=0.73, p=0.01$ and $t=-2.81, p=0.02$). The mean reaction time of all correct responses in condition 4 was significantly less than condition 3 (t=-3.69, p$<$0.01).

\begin{table}[H]
    \centering
    \caption{The summary of significant results in performance, situation awareness, workload, and work memory. Values are presented as median ± standard deviation. Bold values indicate significant results with a p-value less than 0.05 or less than 0.01. Values in task: 1 means monitoring and check task, 2 means response task, / means average all tasks. }
    \begin{tabular}{p{3cm} p{0.3cm} p{1.5cm} p{1.5cm} p{1.5cm} p{1.5cm} p{1cm} p{1cm} p{1cm}} 
        \toprule
        Feature & Task & C1 & C2 & C3 & C4 & p C2-C1 & p C3-C2 & p C4-C3  \\ 
        \midrule
        \multicolumn{9}{l}{\emph{\textbf{Performance}}} \\ 
        error counts& 2 & 4.00±6.91 & 3.41±5.65 & 4.78±6.95 & 2.89±2.47 & 0.21 & \textbf{0.02} & 0.10  \\ 
        fault detection time & 1 & 3.71±2.20 & 2.82±1.13 & 3.03±1.41 & 3.14±2.12 & \textbf{0.05} & 0.37 & 0.27  \\ 
        fault confirm time & 1 & 33.33±14.35 & 28.65±12.65 & 26.67±10.99 & 25.35±8.68 & \textbf{$<$0.01} & 0.18 & 0.91  \\ 
        completion time & / & 82.76±37.41 & 73.64±31.12 & 70.91±23.18 & 61.54±18.38 & \textbf{0.05} & 0.37 & \textbf{$<$0.01}  \\ 
        \multicolumn{9}{l}{\emph{\textbf{Situation awareness}}} \\
        Total score & 1 & 82.59±21.14 & 81.85±24.07 & 86.30±18.94 & 80.37±26.31 & 0.82 & 0.18 & \textbf{0.03}  \\ 
        comprehension  & 2 & 23.33±4.16 & 23.89±3.49 & 22.96±3.74 & 23.70±2.97 & 0.08 & \textbf{0.02} & 0.16  \\ 
        communication  & 1 & 4.44±1.60 & 4.07±1.98 & 4.63±1.33 & 3.89±2.12 & 0.33 & 0.08 & \textbf{0.04}  \\ 
        \multicolumn{9}{l}{\emph{\textbf{Workload}}} \\ 
        physical demand  & / & 23.68±19.11 & 24.18±17.50 & 30.17±20.14 & 43.56±22.81 & 0.56 & \textbf{0.01} & \textbf{$<$0.01}  \\ 
        temporal demand  & / & 43.74±22.57 & 38.26±19.44 & 38.37±22.16 & 41.73±22.06 & \textbf{0.03} & 0.97 & 0.16  \\ 
        effort & / & 44.63±23.86 & 39.08±22.10 & 40.93±22.00 & 46.94±22.32 & 0.09 & 0.33 & \textbf{0.03}  \\ 
        total score & / & 48.40±29.30 & 42.78±22.55 & 43.42±23.50 & 46.47±22.35 & \textbf{0.03} & 0.72 & 0.09  \\ 
        \multicolumn{9}{l}{\emph{\textbf{Work memory}}} \\
        missed number & / & 0.62±0.90 & 0.27±0.53 & 0.73±0.96 & 0.92±2.56 & \textbf{0.01} & \textbf{0.02} & 0.73  \\ 
        mean reaction time of correct responses & / & 1.12±0.18 & 1.10±0.16 & 1.07±0.17 & 1.02±0.16 & 0.39 & 0.19 & \textbf{$<$0.01}  \\ 
        \bottomrule
    \end{tabular}
    \label{table:task result}
\end{table}

\subsection{Physiological measurement results}
\subsubsection{ECG and HRV}
\begin{figure}[H]
    \centering
    \includegraphics[width=1\linewidth]{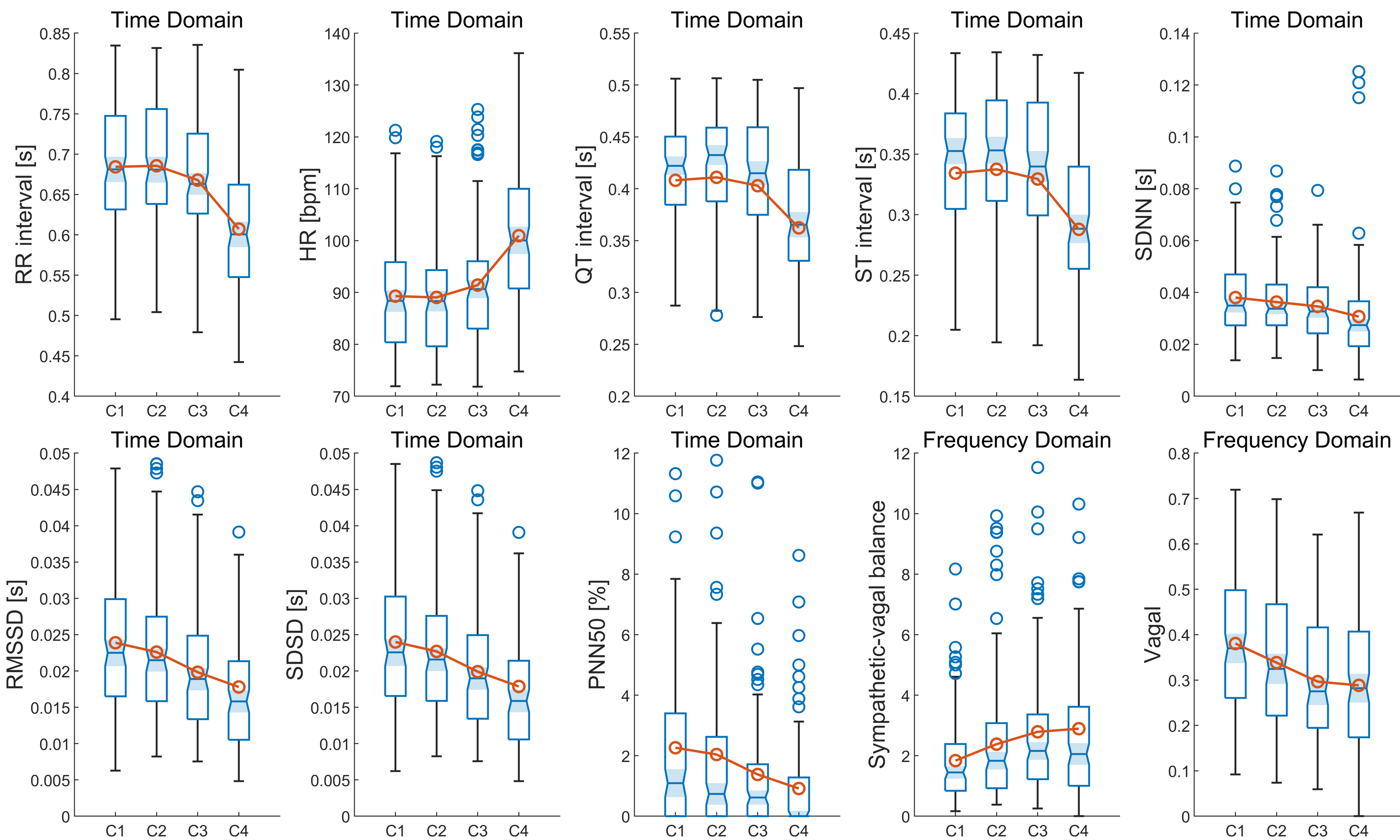}
    \caption{The boxplots display significant results in ECG features. Red lines represent average values.}
    \label{fig:ecg result}
\end{figure}

ANOVA was applied among all the ECG features across adjacent conditions. All the significant p values are shown in \Cref{table:ecg result} and visualized as boxplots with average lines in \Cref{fig:ecg result}. Statistical significance is accepted at $p<0.05$. \par
\textit{\textbf{Time domain}}. RR interval and ST interval in condition 4 were greatly less than condition 1-3 ($F=27.64, p<0.01$ and $F=4.28, p=0.02, 0.01, 0.04$). And QT interval in condition 4 was greatly shorter than condition 1-2 ($F=3.86, p=0.03, 0.02$). HR in condition 4 was greatly faster than condition 1-3 ($F=29.46, p<0.01$). Furthermore, SDNN, RMSSD, SDSD, and PNN50 in condition 4 were greatly less than condition 1-2 ($F=5.76, p\le 0.01$ and $F=8.37, p<0.01$ and $F=8.34, p<0.01$ and $F=7.96, p<0.01$). RMSSD, SDSD, PNN50 in condition 3 were significantly less than condition 1 ($F=8.37, p=0.01$ and $F=8.34, p=0.01$ and $F=7.96, p=0.02$). \par
\textit{\textbf{Frequency domain}}. LF/HF in condition 3-4 was greatly more than condition 1 ($F=5.45, p\le 0.01$). Vagal in condition 3-4 was greatly less than condition 1 ($F=6.44, p<0.01$). \par
\begin{figure}[H]
    \centering
    \includegraphics[width=0.8\linewidth]{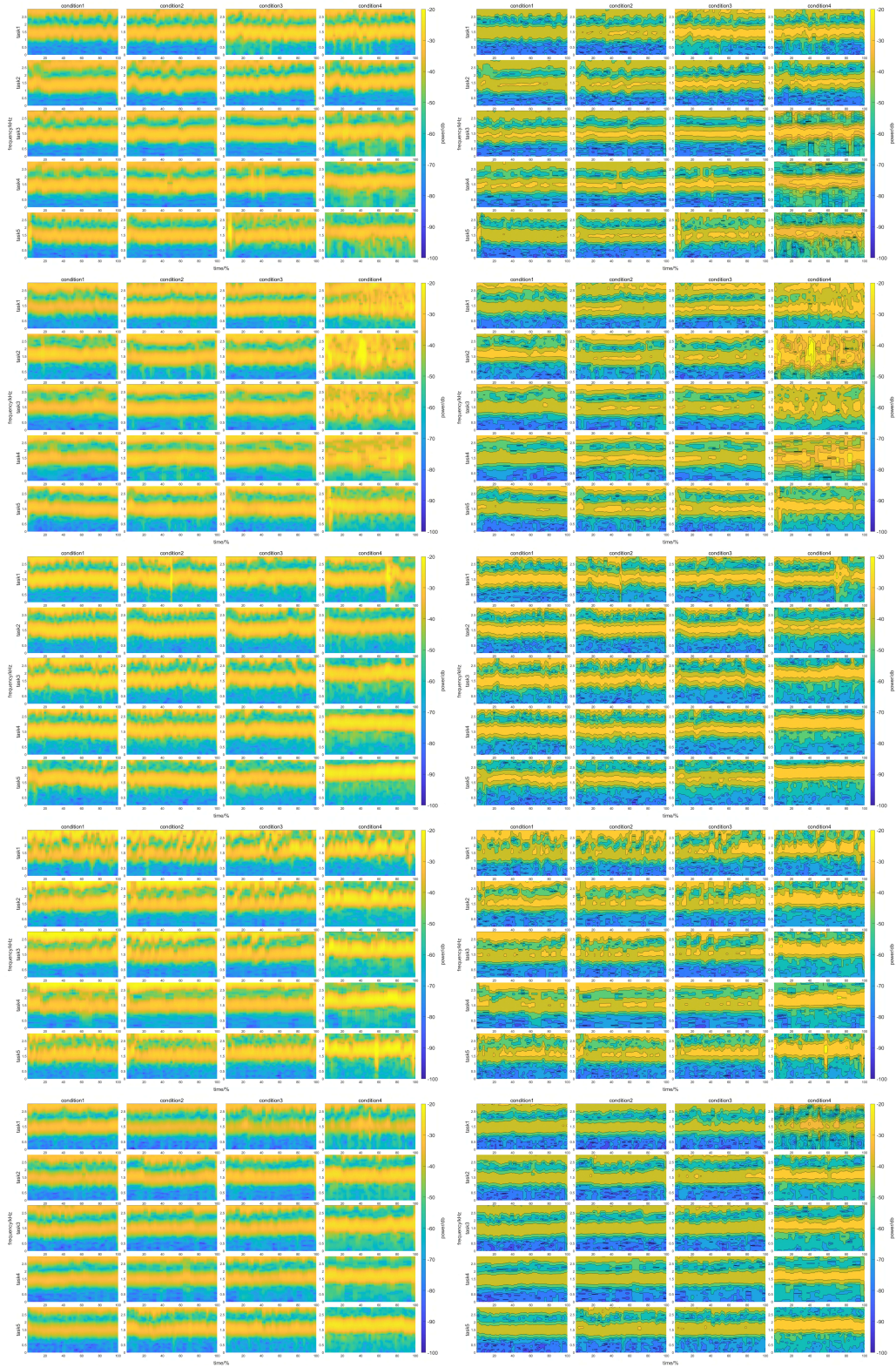}
    \caption{The typical time-frequency representation of RR time series for 9 subjects among different conditions (in columns) and different tasks (in rows). left plots are the raw time-frequency representation of the RR time series and right plots with contour lines on.}
    \label{fig:ecg tf all}
\end{figure}

\textit{\textbf{Time-frequency domain}}. As shown in \Cref{fig:ecg tf all}, a frequency band could be found around 1.5 Hz. PSD of the frequency band increased among different tasks and decreased in the last task. Band frequency, high-frequency component, and overall PSD increased with temperature and humidity.

\begin{table}[H]
    \centering
    \caption{The summary of significant results in ECG features. Values are presented as median ± standard deviation. Bold values indicate significant results with a p-value less than 0.05.}
    \begin{tabular}{p{1.7cm} p{1.2cm} p{1.2cm} p{1.2cm} p{1.2cm} p{1cm} p{1cm} p{1cm} p{1cm} p{1cm} p{1cm}}
        \toprule
        Feature & C1 & C2 & C3 & C4 & p C1-C2 & p C1-C3 & p C1-C4 & p C2-C3 & p C2-C4 & p C3-C4  \\ 
        \midrule
        \multicolumn{11}{l}{\emph{\textbf{Time domain}}} \\ 
        RR interval [s] & 0.68±0.08 & 0.69±0.08 & 0.67±0.08 & 0.61±0.08 & 1.00 & 0.34 & \textbf{$<$}\textbf{0.01} & 0.27 & $<$\textbf{0.01} & $<$\textbf{0.01}  \\ 
        HR [bpm] & 89.29±11.08 & 89.05±10.91 & 91.42±11.32 & 100.94±13.77 & 1.00 & 0.45 & $<$\textbf{0.01} & 0.35 & $<$\textbf{0.01} & $<$\textbf{0.01}  \\ 
        QT interval [s] & 0.41±0.13 & 0.41±0.13 & 0.40±0.13 & 0.36±0.13 & 1.00 & 0.99 & \textbf{0.03} & 0.96 & \textbf{0.02} & 0.06  \\ 
        ST interva [s] & 0.33±0.13 & 0.34±0.13 & 0.33±0.13 & 0.29±0.13 & 1.00 & 0.99 & \textbf{0.02} & 0.96 & \textbf{0.01} & \textbf{0.04}  \\ 
        SDNN [s] & 0.04±0.01 & 0.04±0.01 & 0.03±0.01 & 0.03±0.02 & 0.78 & 0.25 & $<$\textbf{0.01} & 0.80 & \textbf{0.01} & 0.14  \\ 
        RMSSD [s] & 0.02±0.01 & 0.02±0.01 & 0.02±0.01 & 0.02±0.02 & 0.77 & \textbf{0.01} & $<$\textbf{0.01} & 0.16 & $<$\textbf{0.01} & 0.43  \\ 
        SDSD [s] & 0.02±0.01 & 0.02±0.01 & 0.02±0.01 & 0.02±0.02 & 0.77 & \textbf{0.01} & $<$\textbf{0.01} & 0.16 & $<$\textbf{0.01} & 0.43  \\ 
        PNN50 [\%] & 2.26±3.08 & 2.04±2.97 & 1.38±2.12 & 0.92±1.47 & 0.88 & \textbf{0.02} & $<$\textbf{0.01} & 0.14 & $<$\textbf{0.01} & 0.44  \\ 
        \multicolumn{11}{l}{\emph{\textbf{Frequency domain}}} \\ 
        Sympathetic-vagal balance & 1.84±1.41 & 2.38±2.08 & 2.79±2.36 & 2.89±3.30 & 0.24 & \textbf{0.01} & $<$\textbf{0.01} & 0.50 & 0.29 & 0.98  \\ 
        Vagal & 0.38±0.20 & 0.34±0.19 & 0.30±0.18 & 0.29±0.20 & 0.28 & $<$\textbf{0.01} & $<$\textbf{0.01} & 0.29 & 0.15 & 0.99  \\ 
        \bottomrule
    \end{tabular}
    \label{table:ecg result}
\end{table}

\subsubsection{fNIRS}
\begin{figure}[H]
    \centering
    \includegraphics[width=0.8\linewidth]{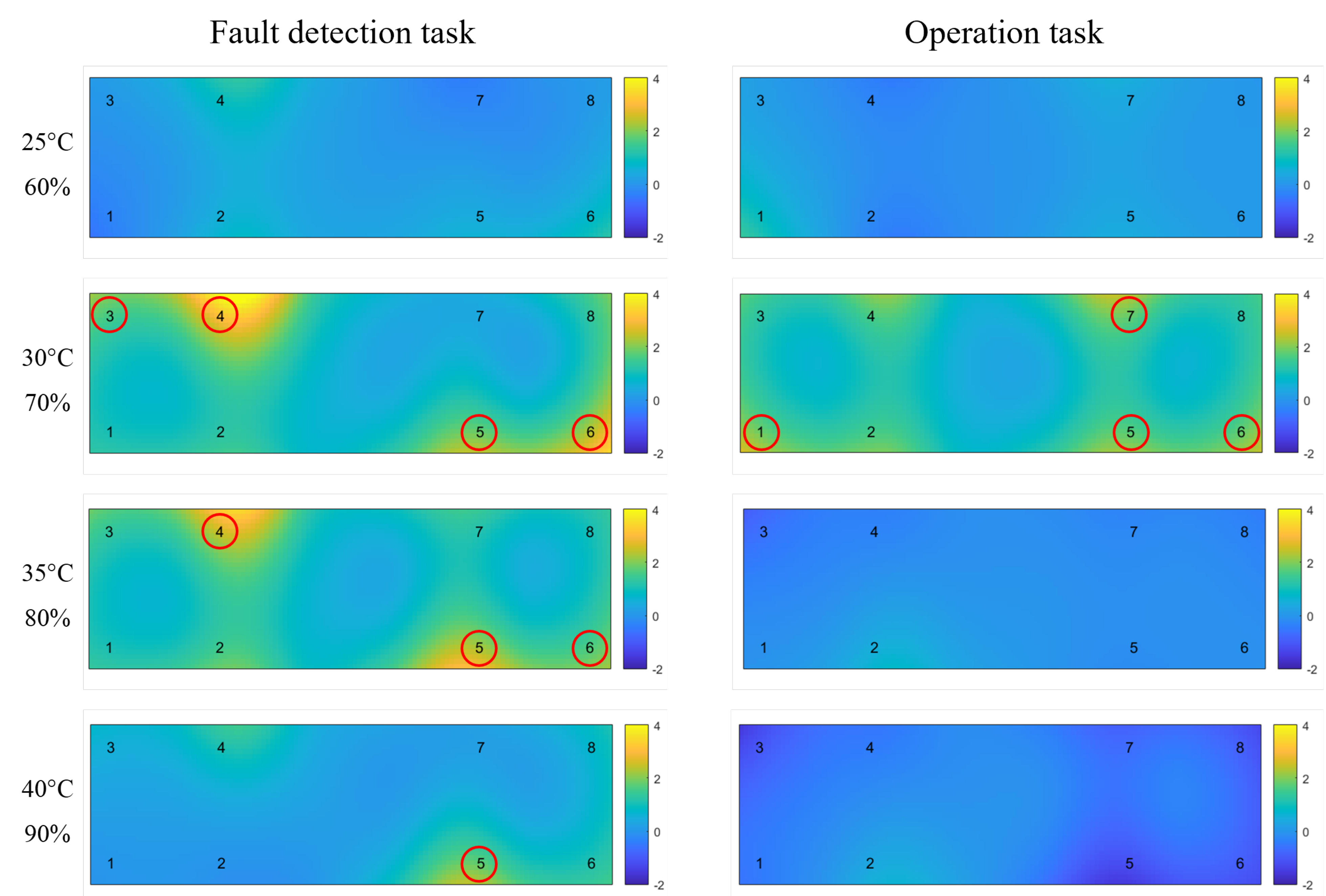}
    \caption{The PFC activation map among conditions and tasks. Red circles reveal valid activated channels (betas differ from 0)}
    \label{fig:fnirs beta}
\end{figure}

\textit{\textbf{GLM results}}. One sample t-test (differ from 0) was applied among GLM betas for tasks and conditions. All the significant p values are shown in \Cref{table:fnirs beta} and visualized as a PFC activation map in \Cref{fig:fnirs beta}. Statistical significance is accepted at p $<$ 0.05. Results revealed that there was no activated channel in condition 1. In condition 2, there were 4 activated channels (ch3, ch4, ch5, ch6) in the fault detection task and 4 activated channels (ch1, ch5, ch6, ch7) in the operation task. In condition 3, there were 3 activated channels (ch4, ch5,ch6) in the fault detection task and no activated channels in the operation task. In condition 4, there was only 1 activated channel (ch5) in the fault detection task and 0 in the operation task.

\begin{table}[H]
    \centering
    \caption{The t values for GLM betas. Bold values indicate significant results with a p-value less than 0.05. The detection task had more activated channels than the operation task. Condition 2 and 3 had more activated channels.}
    \begin{tabular}{ccccccccccc}
    %{p{2cm} p{2cm} p{1cm} p{1cm} p{1cm} p{1cm} p{1cm} p{1cm} p{1cm} p{1cm} p{1cm}}
        \toprule
        condition & task & Valid channels & ch1 p & ch2 p & ch 3 p & ch 4 p & ch5 p & ch6 p & ch7 p & ch8 p  \\ 
        \midrule
        \multirow[t]{2}{*}{25°C 60\%} 
         & detection & 0  & 0.59  & 0.45  & 0.94  & 0.21  & 0.57  & 0.24  & 0.69  & 0.89   \\ 
        ~ & operation & 0  & 0.19  & 0.69  & 0.83  & 0.66  & 0.65  & 0.93  & 0.62  & 0.97   \\ 
        \multirow[t]{2}{*}{30°C 70\%}
        & detection & 4  & 0.21  & 0.15  & \textbf{0.04}  & \textbf{0.00}  & \textbf{0.02}  & \textbf{0.00}  & 0.69  & 0.06   \\ 
        ~ & operation & 4  & \textbf{0.02}  & 0.07  & 0.10  & 0.05  & \textbf{0.04}  & \textbf{0.02}  & \textbf{0.02}  & 0.13   \\ 
        \multirow[t]{2}{*}{35°C 80\%} 
        & detection & 3  & 0.19  & 0.07  & 0.07  & \textbf{0.00}  & \textbf{0.00}  & \textbf{0.03}  & 0.15  & 0.18   \\ 
        ~ & operation & 0  & 0.96  & 0.46  & 0.38  & 0.72  & 0.88  & 0.94  & 0.76  & 0.66   \\ 
        \multirow[t]{2}{*}{40°C 90\%}
        & detection & 1  & 0.97  & 0.92  & 0.46  & 0.15  & \textbf{0.02}  & 0.13  & 0.86  & 0.28   \\ 
        ~ & operation & 0  & 0.57  & 0.55  & 0.15  & 0.63  & 0.14  & 0.25  & 0.40  & 0.22   \\ \hline
    \end{tabular}
    \label{table:fnirs beta}
\end{table}

\begin{figure}[H]
    \centering
    \includegraphics[width=1\linewidth]{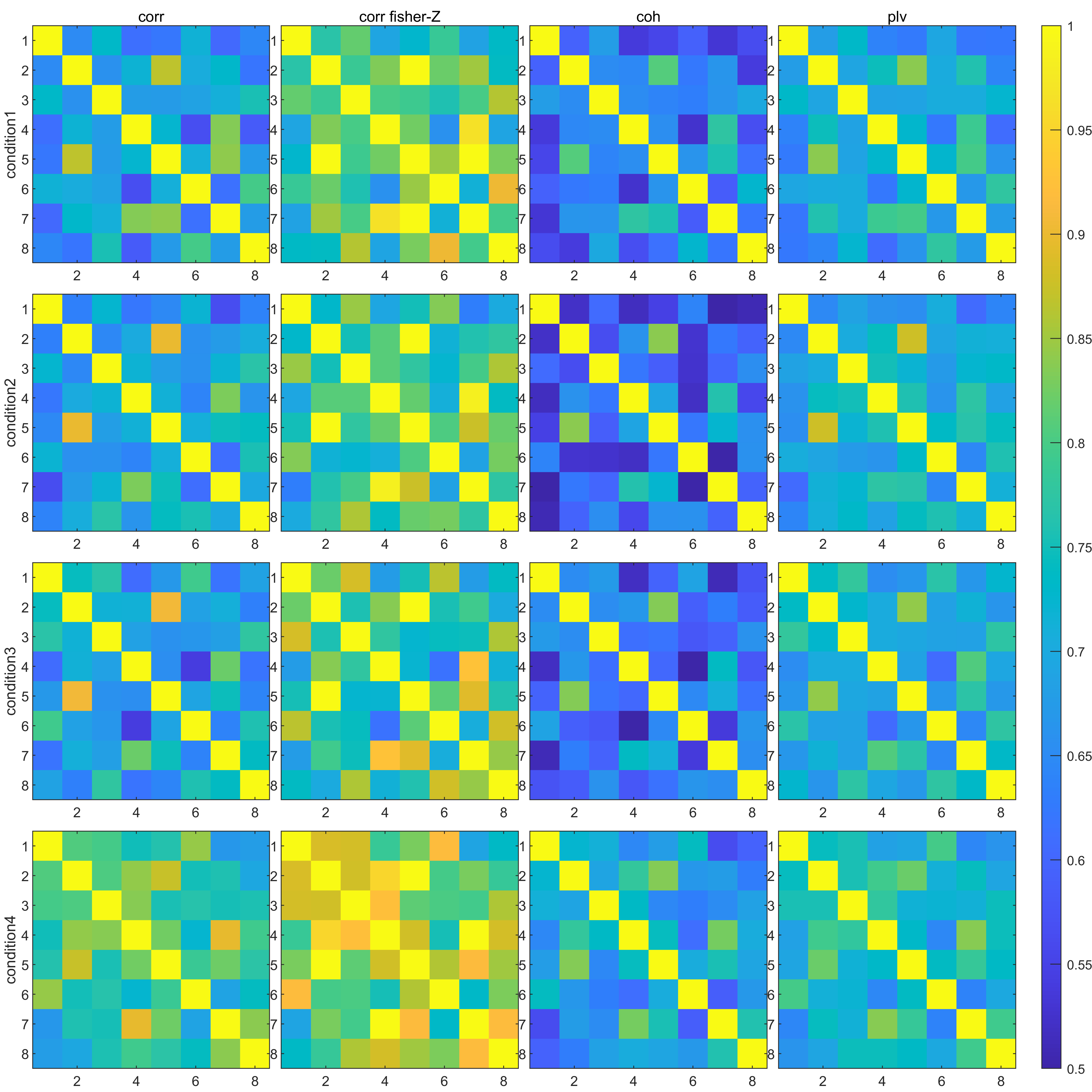}
    \caption{Four kings of FC matrix (CPRR, CORR fisher-Z, COH, PLV) for 4 conditions.}
    \label{fig:fnirs fc all}
\end{figure}

\begin{figure}[H]
    \centering
    \includegraphics[width=0.5\linewidth]{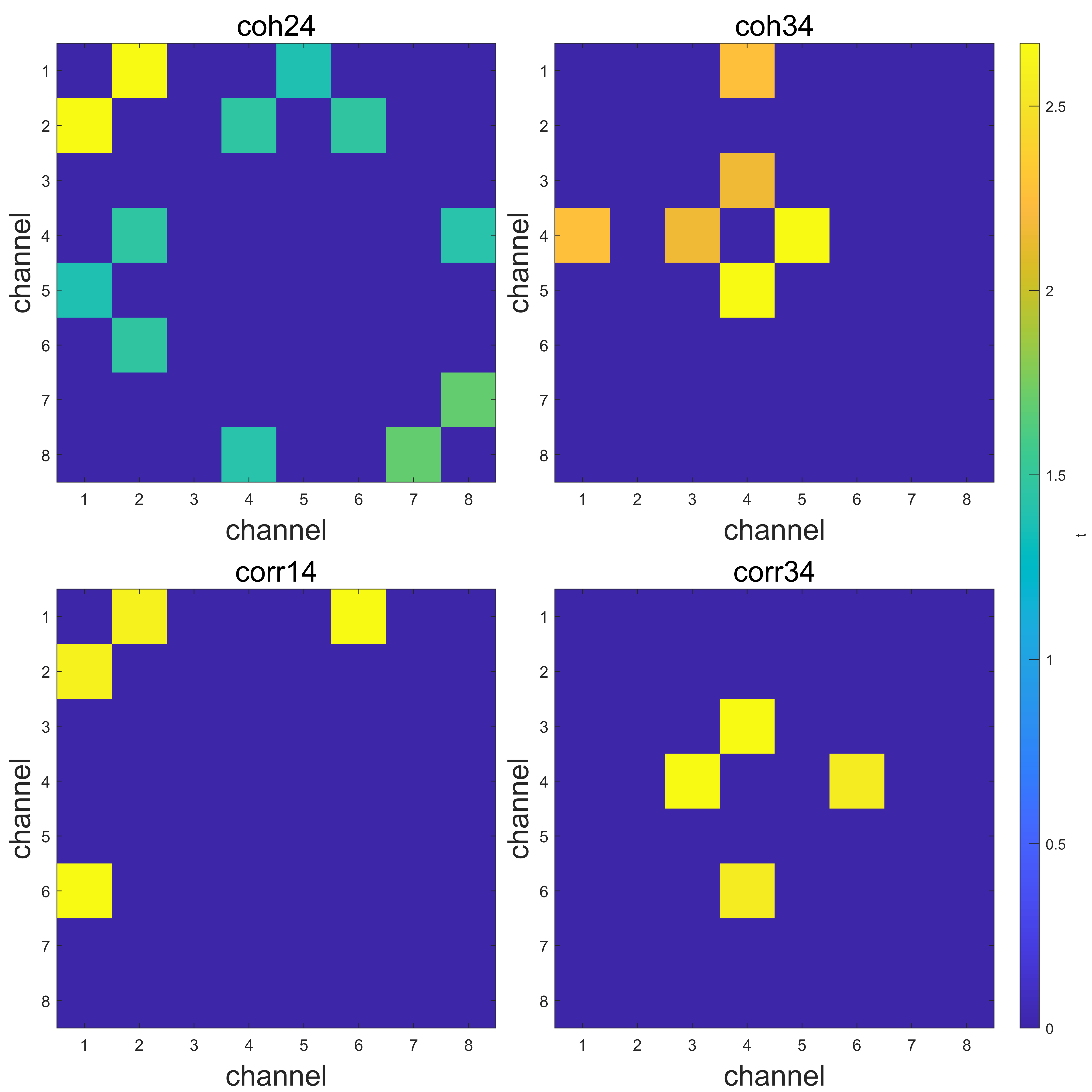}
    \caption{Valid FC matrix after NBS correction}
    \label{fig:fnirs fc sig}
\end{figure}

\textit{\textbf{Resting-state FC results}}. Four kinds of FC matrix (CORR, CORR fisher-Z, COH, PLV) were plotted in \Cref{fig:fnirs fc all}. NBS correction (multiple comparisons correction) was applied and valid results are listed in \Cref{table:fnirs fc sig} and plotted in \Cref{fig:fnirs fc sig}. For coherence (COH), frequency synchronization, there were great differences between condition 2 and 4 ($p<0.01$) in ch1 and ch2 ($t=5.08$), ch1 and ch5 ($t=2.63$), ch2 and ch4 ($t=2.79$), ch2 and ch6 ($t=2.80$), ch4 and ch8 ($t=2.71$), ch7 and ch8 ($t=3.23$) and between condition 3 and 4 ($p=0.03$) in ch1 and ch4 ($t=2.81$), ch3 and ch4 ($t=2.69$), ch4 and ch5 ($t=3.33$). For correlation (CORR), the time synchronization, there were great differences between condition 1 and 4 ($p=0.04$) in ch1 and ch2 ($t=2.93$), ch1 and ch6 ($t=3.00$), and between condition 3 and 4 ($p=0.05$) in ch3 and ch4 ($t=2.67$), ch4 and ch6 ($t=2.56$). No significant difference was reported among any condition and any channel for PLV.

\begin{table}[H]
    \centering
    \caption{Valid results after NBS correction}
    \begin{tabular}{lllllll}
        \toprule
        feature & Condition1 & Condition 2 & Channel 1 & Channel 2 & t & p  \\ 
        \midrule
        \multirow[t]{9}{*}{Coherence} &
            \multirow[t]{6}{*}{2} & \multirow[t]{6}{*}{4}
              & 1 & 2 & 5.08 & \multirow[t]{6}{*}{$<0.01$}  \\ 
            ~ & ~ & ~ & 1 & 5 & 2.63 & ~  \\ 
            ~ & ~ & ~ & 2 & 4 & 2.79 & ~  \\ 
            ~ & ~ & ~ & 2 & 6 & 2.80 & ~  \\ 
            ~ & ~ & ~ & 4 & 8 & 2.71 & ~  \\ 
            ~ & ~ & ~ & 7 & 8 & 3.23 & ~  \\
            ~ & \multirow[t]{3}{*}{3} & \multirow[t]{3}{*}{4} 
              & 1 & 4 & 2.81 & \multirow[t]{3}{*}{0.03}  \\ 
            ~ & ~ & ~ & 3 & 4 & 2.69 & ~  \\ 
            ~ & ~ & ~ & 4 & 5 & 3.33 & ~  \\ 
        \multirow[t]{4}{*}{Correlation} &
            \multirow[t]{2}{*}{1} & \multirow[t]{2}{*}{4}
              & 1 & 2 & 2.93 & \multirow[t]{2}{*}{0.04}  \\ 
            ~ & ~ & ~ & 1 & 6 & 3.00 & ~  \\ 
            ~ & \multirow[t]{2}{*}{3} & \multirow[t]{2}{*}{4}
              & 3 & 4 & 2.67 & \multirow[t]{2}{*}{0.05}  \\ 
            ~ & ~ & ~ & 4 & 6 & 2.56 & ~  \\ 
        \bottomrule
    \end{tabular}
    \label{table:fnirs fc sig}
\end{table}

\subsubsection{Eye tracking}
\begin{figure}[H]
    \centering
    \includegraphics[width=1\linewidth]{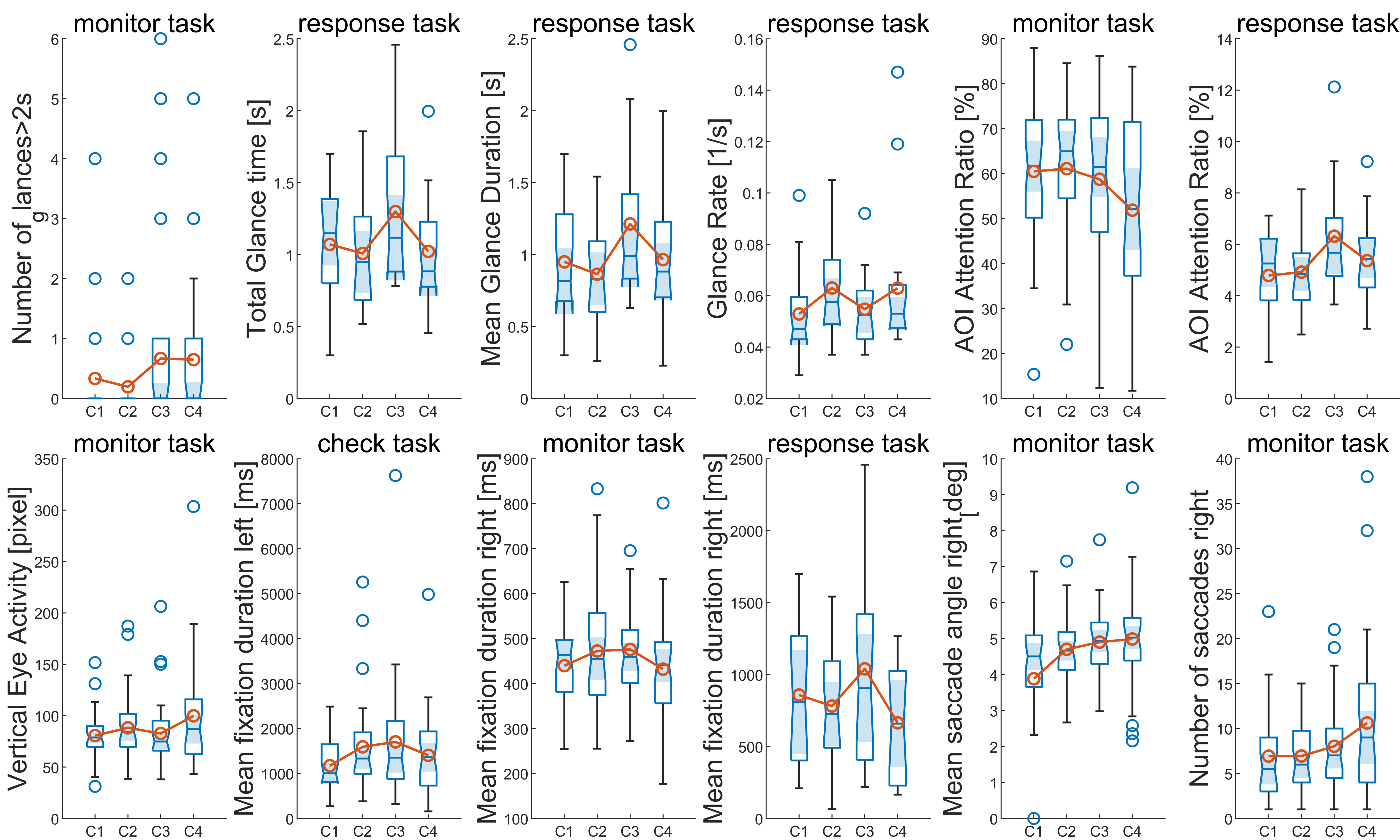}
    \caption{The boxplots display significant results in AOI-related features. Red lines represent average values.}
    \label{fig:aoi result}
\end{figure}

\begin{figure}[H]
    \centering
    \includegraphics[width=1\linewidth]{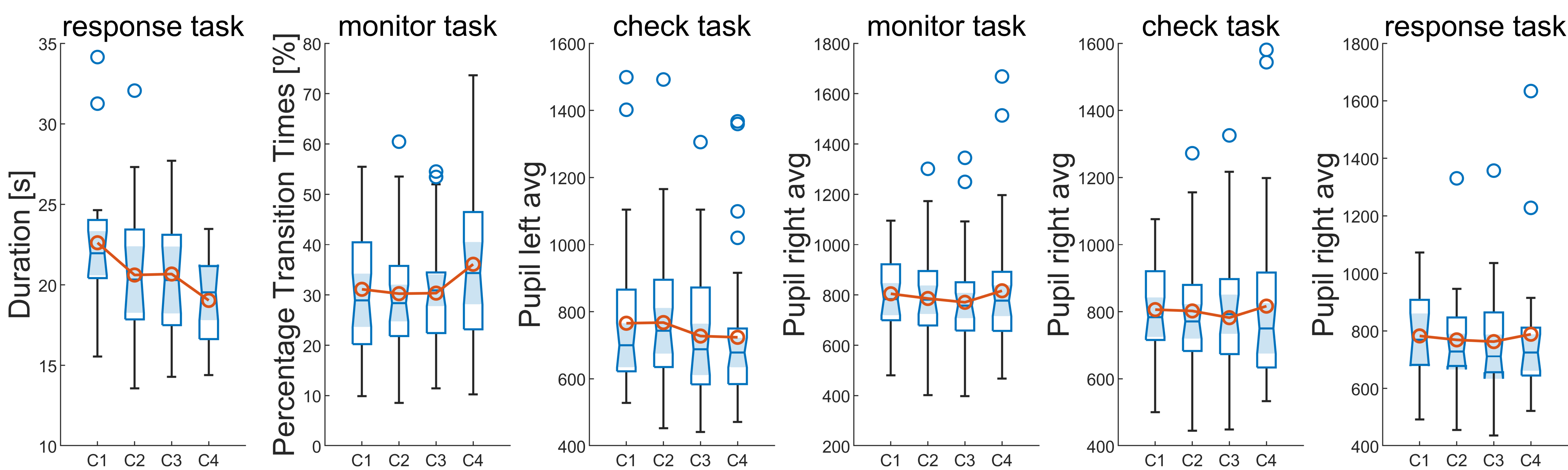}
    \caption{The boxplots display significant results in pupil-related features. Red lines represent average values.}
    \label{fig:pp result}
\end{figure}

One sample t-test (differ from 0) was applied among all the eye-tracking features across adjacent conditions. All the significant p values are shown in \Cref{table:aoi result} and \Cref{table:pp result} and visualized as boxplots with average lines in \Cref{fig:aoi result} and \Cref{fig:pp result}. Statistical significance is accepted at $p<0.05$. \par
\textit{\textbf{AOI results}}. As shown in \Cref{table:aoi result} and \Cref{fig:aoi result}, the number of single glance more than 2 seconds in condition 3 was greatly more than condition 2 ($t=2.18, p=0.04$). Total glance time, mean glance duration, and AOI attention ratio in condition 3 were significantly more than condition 2 ($t=2.21, p=0.04$ and $t=3.07, p=0.01$ and $t=2.58, p=0.02$) and more than condition 4 ($t=-2.57 p=0.02$ and $t=-2.28, p=0.04$ and $t==3.23, p<0.01$). Glance rate, mean fixation duration and mean saccade angle right in condition 2 were greatly more than condition 1 ($t=2.63, p=0.02$ and $t=2.14, p=0.04$ and $t=2.41, p=0.02$). Vertical eye activity and the number of saccades right in condition 4 were more than condition 3 ($t=2.73, p=0.01$ and $t=2.05, p=0.05$). The mean fixation duration right in condition 4 was greatly less than condition 3 ($t=-2.27, p=0.03$).

\begin{table}[H]
    \centering
    \caption{The summary of significant results in AOI-related features. Values are presented as median ± standard deviation. Bold values indicate significant results with a p-value less than 0.05. Values in task: 1 means monitoring task, 2 means check task, 3 means response task.}
    \begin{tabular}{p{3.7cm} p{0.3cm} p{1.8cm} p{1.8cm} p{1.8cm} p{1.8cm} p{1cm} p{1cm} p{1cm}}
        \toprule
        Feature & Task & C1 & C2 & C3 & C4 & p C2-C1 & p C3-C2 & p C4-C3  \\ 
        \midrule
        Number of glances$>$2s & 1 & 0.33±0.79 & 0.19±0.47 & 0.67±1.47 & 0.65±1.30 & 0.36 & \textbf{0.04} & 0.76  \\ 
        Total Glance time [s] & 3 & 1.07±0.40 & 1.01±0.39 & 1.30±0.52 & 1.02±0.38 & 0.46 & \textbf{0.04} & \textbf{0.02}  \\ 
        Mean Glance Duration [s] & 3 & 0.95±0.39 & 0.86±0.38 & 1.21±0.53 & 0.96±0.43 & 0.46 & \textbf{0.01} & \textbf{0.04}  \\ 
        Glance Rate [1/s] & 3 & 0.05±0.02 & 0.06±0.02 & 0.05±0.01 & 0.06±0.03 & \textbf{0.02} & 0.09 & 0.28  \\ 
        AOI Attention Ratio [\%] & 1 & 60.50±15.31 & 61.10±15.95 & 58.75±18.20 & 51.87±22.15 & 0.82 & 0.31 & $<$\textbf{0.01}  \\ 
        AOI Attention Ratio [\%] & 3 & 4.79±1.66 & 4.92±1.57 & 6.32±2.26 & 5.36±1.71 & 0.72 & \textbf{0.02} & 0.08  \\ 
        Vertical Eye Activity [pixel] & 1 & 80.66±22.55 & 88.18±32.22 & 82.66±33.15 & 99.87±51.64 & 0.16 & 0.18 & \textbf{0.01}  \\ 
        Mean fixation duration left [ms] & 2 & 1172.75±565.56 & 1592.61±1014.11 & 1702.61±1314.78 & 1404.03±929.85 & \textbf{0.05} & 0.72 & 0.27  \\ 
        Mean fixation duration right [ms] & 1 & 439.73±97.46 & 472.45±131.91 & 475.96±96.15 & 432.08±121.18 & \textbf{0.04} & 0.88 & \textbf{0.03}  \\ 
        Mean fixation duration right [ms] & 3 & 857.67±484.26 & 781.05±430.88 & 1039.89±676.90 & 664.45±392.81 & 0.99 & 0.08 & \textbf{0.04}  \\ 
        Mean saccade angle right [deg] & 1 & 3.88±1.97 & 4.70±0.94 & 4.90±0.95 & 4.99±1.38 & \textbf{0.02} & 0.32 & 0.79  \\ 
        Number of saccades right & 1 & 6.93±5.17 & 6.94±3.60 & 8.03±5.40 & 10.62±8.37 & 0.67 & 0.30 & \textbf{0.05}  \\ 
        \bottomrule
    \end{tabular}
    \label{table:aoi result}
\end{table}

\textit{\textbf{Pupil results}}. As shown in \Cref{table:pp result} and \Cref{fig:pp result}, duration in condition 4 was greatly less than condition 3 ($t=-2.21, p=0.04$) and condition 2 less than condition 1 ($-2.48, p=0.02$). Percentage transition times in condition 3 was greatly less than condition 4 ($t=2.39, p=0.02$). Pupil diameter decreased from condition 1-3 ($t=-2.93, p=0.01$ and $t=-3.29, p<0.01$) and increased from condition 3-4 ($t=2.59, p=0.01$). 

\begin{table}[H]
    \centering
    \caption{The summary of significant results in pupil-related features. Values are presented as median ± standard deviation. Bold values indicate significant results with a p-value less than 0.05. Values in task: 1 means monitoring task, 2 means check task, 3 means response task.}
    \begin{tabular}{p{2.5cm} p{0.5cm} p{1.8cm} p{1.8cm} p{1.8cm} p{1.8cm} p{1cm} p{1cm} p{1cm} }
        \toprule
        Feature & Task & C1 & C2 & C3 & C4 & p C2-C1 & p C3-C2 & p C4-C3 \\ 
        \midrule
        Duration [s] & 3 & 22.62±4.56 & 20.61±4.73 & 20.67±4.06 & 19.03±2.74 & \textbf{0.02} & 0.94 & \textbf{0.04} \\ 
        Percentage Transition Times [\%] & 1 & 31.12±12.07 & 30.25±11.89 & 30.36±11.43 & 36.10±16.72 & 0.71 & 0.95 & \textbf{0.02} \\ 
        Pupil left avg & 2 & 765.62±227.20 & 767.62±214.42 & 727.42±200.13 & 723.69±218.04 & 0.86 & $<\textbf{0.01}$ & 0.98 \\ 
        Pupil right avg & 1 & 804.64±154.96 & 785.83±179.94 & 770.75±199.82 & 816.48±258.55 & \textbf{0.01} & 0.27 & \textbf{0.01} \\ 
        Pupil right avg & 2 & 806.17±145.28 & 802.25±179.40 & 782.20±185.46 & 816.61±252.90 & \textbf{0.02} & 0.14 & \textbf{0.05} \\ 
        Pupil right avg & 3 & 782.45±147.29 & 768.69±183.24 & 762.66±209.55 & 788.31±275.54 & \textbf{0.02} & 0.70 & 0.26 \\ 
        \bottomrule
    \end{tabular}
    \label{table:pp result}
\end{table}

\begin{figure}[H]
    \centering
    \includegraphics[width=1\linewidth]{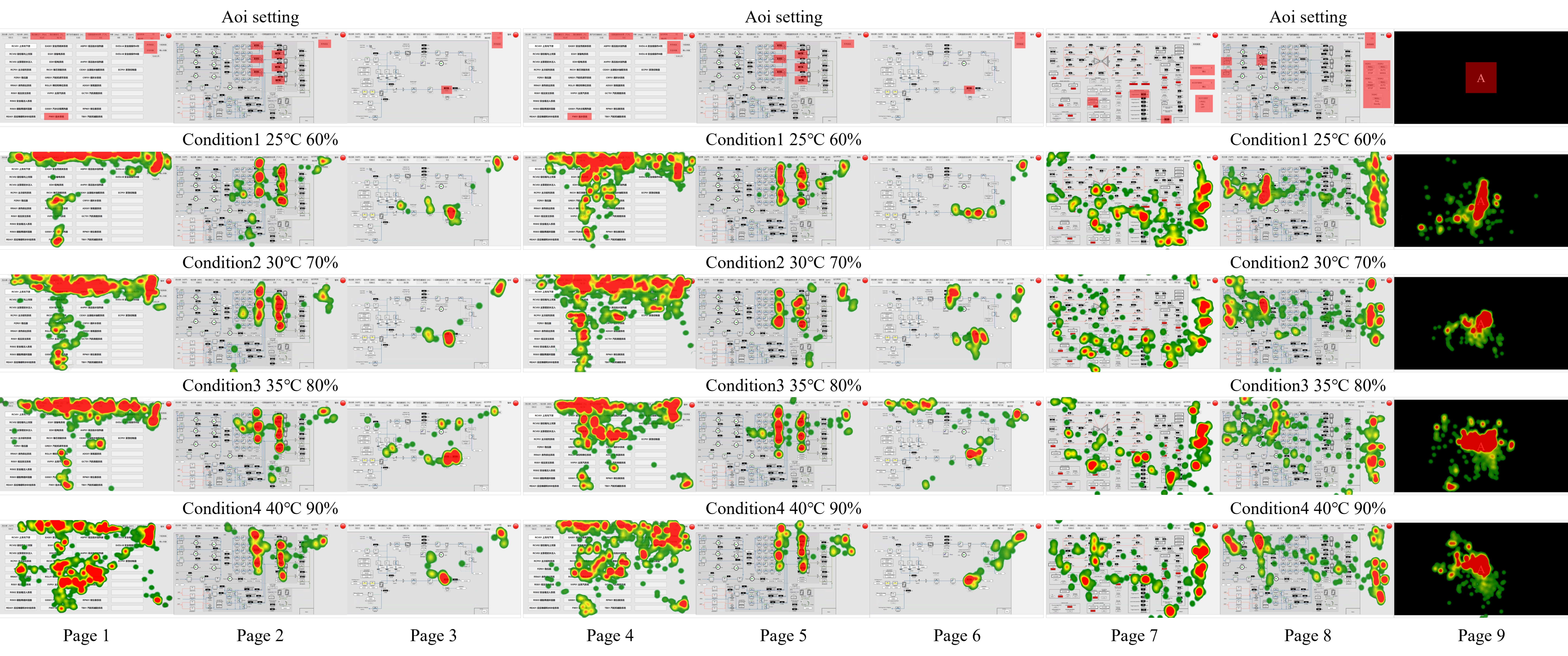}
    \caption{The heat map. Row 1: AOI settings in every interest page. Row 2-5: condition 1-4. Column 1-3: interest pages in the major task 1: fault detection task. Column 4-6: interest pages in the major task 2: fault detection task. Column 7-8: interest pages in the major task 3: operation task. Column 9: 2-back task.}
    \label{fig:heating map}
\end{figure}

\textit{\textbf{Heat map}}. As shown in \Cref{fig:heating map}, the heat map revealed attention distribution on each interest page. In condition 1 (row 2), heat distribution was wider to gather more information. In condition 2 (row 3), distribution was more precise and focused on AOIs. In condition 3 (row 4), areas near AOIs and confusing information got more fixation. In condition 4 (row 5), glances were messy on interest pages. Furthermore, within the procedure of experiment in every condition (column 1 and column 4, 2 and 5, 3 and 6), distribution was messier as the result of workload accumulation in the hot and humid environment. 

%% file: files/05discussion.tex
\section{Discussion}
\subsection{The effect of hot and humid environment on task performance and cognitive functions}
Among all conditions, no difference was observed in the task completion rate, SA level 1 perception score, and SA level 3 projection score. This suggests that high temperature and humidity had little impact on the completion rate. However, they did have a significant influence on the time and accuracy of the NPPs task. \par
In condition 1, longer fault detection time and fault confirmation time with higher temporal demand were observed. The result revealed that the performance would decline slightly at low temperature and humidity, which was also proved by Pilcher \citep{5pilcher2002effects}.  He explored the threshold of the maximal adaptability model by meta-analysis and found that there was a negative influence on reasoning, learning, and memory tasks under 21.11℃ WBGT ( condition 1 was 21.26℃ WBGT in our study). Our studies suggest that this is caused by nervous mode at the beginning of an emergency.\par
In condition 2, shorter fault detection time, fault confirmation time, and task completion time along with lower temporal demand as well as the lowest task load index and fewer missed numbers in the 2-back tasks were obtained. These results suggest that an environment in 26.88℃ WBGT was suitable for operators with lower workload, calmer disposition, and better working memory, ultimately leading to better performance. The finding was supported by the inverse U-shaped model and the maximal adaptability model. Furthermore, Pilcher \citep{5pilcher2002effects} found a small improvement in reasoning, learning, and memory tasks when exposed to a hot environment of 26.67℃ WBGT or above.  \par
In condition 3, there were more errors (clicking on the wrong place or making incorrect decisions) in the operation task and a lower SA level 2 comprehension score which meant more mistakes and weaker comprehensive ability leading to poorer performance. The decrease was predicted by the maximal adaptability model and has been mentioned by several studies. Ramsey \citep{3ramsey1995} found a significant performance decrement in perceptual-motor tasks (for example, alerting and operating equipment) between 30℃ WBGT and 33℃ WBGT (condition 3 was 31.92℃ WBGT in our study). Pilcher \citep{5pilcher2002effects} mentioned a great performance decline beyond 32.22℃ WBGT. \par
In condition 4 (36.81℃ WBGT), shorter task completion time, and shorter mean reaction time of all correct responses were presented which demonstrated faster response in simple reaction tasks because of stress response under extreme heat stress. Additionally, the most missed numbers in the 2 back task, the worst SAGAT total score, and the worst communication score with higher physical demand and higher effort were presented. The results indicated that extreme heat stress would lead to a rapid decline in complex tasks (i.e. teamwork) and cognitive functions (i.e. situation awareness, working memory) because of the fast accumulation of physical load and mental load, which was opposite to the improvement in the simple task. The similar phenomenon has been observed in numerous studies. Lovingood \citep{2Lov1967} found shorter reaction time, higher classification accuracy, and better calculation ability with worse arm stability under high temperature (52℃ compared to 23.3℃). Hancock \citep{4hancock2003} reviewed that simple performance improved under a 10-minute short-term 43℃ WBGT hot exposure. Tian \citep{7tian2021decreased} studied humidity influence and found that 70\% relative humidity (RH) could lead to worse spatial perception, concentration, working memory, arousal level, and fatigue compared to 50\% RH. Gaoua \citep{56gaoua2011alterations} investigated the heat exposure influence on cognitive functions and found that compared to 20℃, heat discomfort under 50℃ would impair performance on complex cognitive tasks and working memory function. Although the participants in the high-temperature group responded faster, they spent more time finding the right solution \citep{57gaoua2012sensory}. 

\subsection{Physiological-psychological mechanism of impairment on performance and cognitive functions in the hot and humid environment}

\subsubsection{Neural activity based on ECG and HRV}
 Performance decrement around 32℃ WBGT (condition 3) and unsustainable state around 37℃ WBGT (condition 4) could be explained by more tension, worse regulation ability, increasing stress, and mental load reflected by faster HR, lower RMSSD, lower SDSD, lower SDNN, lower PNN50, higher LF/HF and lower vagal. \par
Heart rate (HR), the number of heartbeats per minute, usually increases with heat stress \citep{64ABBASI2020103189} indicating increasing tension, stress, and mental load \citep{23hwang2008predicting, 24de2008cardiovascular}. Heart rate variability (HRV), the relative change in the RR interval between two adjacent heartbeats, generally decreases as a result of stress \citep{22shaffer2014healthy, 23hwang2008predicting} and decreases with increasing task load \citep{24de2008cardiovascular}. The indices describing heart rate variability, such as RMSSD, SDSD, SDNN, and PNN50 \citep{65CHEN2020107372}, often indicate the adaptability of the nervous system to internal and external changes \citep{22shaffer2014healthy}, and generally decrease with the increase of pressure and load \citep{23hwang2008predicting, 28pakarinen2018cardiac}. \par
The high-frequency component of the ECG signal (HF) is primarily associated with the activation of the parasympathetic nervous system \citep{25delaney2000effects}, mainly represented by the vagal nerve. Activation of the vagal nerve, also known as the rest and digest system, is often accompanied by physiological activities such as miosis, heart rate slowing, and bronchoconstriction. The low-frequency component of the ECG signal (LF) mainly reflects the combined activation of the sympathetic and the parasympathetic nervous system \citep{26martelli2014low}, along with physiological activities such as salivation. Therefore, LF/HF is commonly used as an index that reflects the balance between the activation of the sympathetic and parasympathetic nervous systems. It generally increases with pressure \citep{27guasti2005global} and mental load \citep{23hwang2008predicting}. The higher LF/HF under heat stress was also reported in several studies \citep{63LAN201029, 64ABBASI2020103189}.\par
Applying time-frequency analysis, a frequency band of approximately 1.5 Hz was identified. In the same condition, the PSD of the frequency band increased across different tasks associated with accumulating workload. However, there was a decrease in PSD during the final task, suggesting that additional cognitive resources could be collected with effort when participants knew that the procedure was near to end. Moreover, within the same task, the band frequency, high-frequency component, and overall PSD grew
with temperature and humidity. This indicates an accumulating workload and chaotic heart activity.

\subsubsection{Energy metabolism based on fNIRS}
In the study, the detection task had more activated channels (betas significantly differ from 0) than the operation task and more channels were activated in the range of 27-32℃ WBGT (condition 2 and 3). According to the neurovascular coupling mechanism, the neural activity of the brain increases during cognitive activities. As task difficulty \citep{38causse2017mental} or fatigue level \citep{39pan2019applications} increases, there is a regular increment in the activation intensity in the PFC, which is caused by the increasing cognitive resources input and energy consumption of the cognitive functional brain areas. \par
The different performance trends between condition 2 and 3, despite the same PFC activation, could be explained by the neural efficiency hypothesis of intelligent individuals \citep{42neubauer2009intelligence}. The hypothesis claims that more intelligent individuals are able to achieve the same performance while investing fewer cognitive resources and experiencing lower brain activation. With learning and practice, operations might become more automatic and disassociated from PFC activation. Consequently, for more experienced and intelligent individuals, better task performance is associated with lower PFC activation, whereas the opposite holds true for less experienced and intelligent individuals. According to the hypothesis, the activation intensity can reveal the level of mental load \citep{40ayaz2012optical}, but has no significant correlation with task performance \citep{38causse2017mental, 41matsuda2006sustained}. During cognitive tasks in sleep-deprived states, activities in the prefrontal cortex were found to either increase due to compensatory mechanisms or decrease due to cognitive deficits \citep{39pan2019applications}.\par
Therefore, activation in condition 2 indicates an effective cognitive resource input and results in a good performance. In contrast, increased activation in condition 3 was explained by a compensatory mechanism due to fatigue, drowsiness, and slowed thinking. Reduced activation under the highest heat stress conditions (condition 4) indicates inhibition in the prefrontal cortex, an important cortex associated with decision-making, cognitive control, and working memory \citep{fmri2_qian_disrupted_2020}. \par
Applying resting-state functional connectivity analysis, higher synchronization in both time and frequency, which suggests higher cooperation between different brain regions, was observed in the hottest and most humid condition. FC strength is usually associated with fatigue, task complexity, and cognitive control. 
%With the accumulation of fatigue, the FC strength weakens and the coordination of different brain regions decreases, resulting in the decrement of cognitive ability \citep{43xu2017functional}. 
The global functional connectivity strength increases with the task difficulty and cognitive control \citep{44cole2012global}, which was a potential explanation for the fastest response under the heaviest heat stress.

\subsubsection{Visual search pattern based on Eye Tracking}
In the AOI analysis, a greater saccade angle was obtained in condition 2, which meant more accurate search paths associated with a more efficient search strategy under suitable temperature and humidity. \par
In condition 3, we gained a greater number of fixations, longer total glance time, and mean glance duration. Fixation refers to the visual gaze remaining at a particular location, which generally lasts for 200-500 milliseconds. The fixation rate is positively correlated with workload \citep{34wu2020using}. The number of fixations can be used to judge whether individuals need to frequently observe and search on the interface to obtain enough information. Fixation time can be used to estimate the difficulty of extracting information on the page, which has a positive correlation \citep{34wu2020using}.  \par
Meanwhile, we noticed a lower percentage of transition times in condition 3 than in adjacent conditions. Saccades refer to the shifting of the eye between fixations, and generally the frequency of saccades decreases with increasing task difficulty, mental load, and cumulative fatigue \citep{34wu2020using, 36nakayama2002act}.\par
In condition 4, we got more vertical eye activity which indicates worse concentration. Additionally, we observed a shorter mean fixation duration revealing a temporary improvement in cognitive efficiency.\par
The pupil diameter is mainly regulated by internal and external factors. The external factor is light conditions, where darker light leads to a larger pupil diameter. The internal factor is cognitive resources, i.e. mental effort and arousal. Larger pupil diameter is consistently associated with higher cognitive resource investment \citep{30van2018pupil}, mental load, and task complexity \citep{31batmaz2008using}. The average pupil diameter of the right eye in condition 4 was significantly larger than that in condition 3, indicating that the psychological load in condition 4 increased sharply in the extremely high-temperature and high-humidity environment, resulting in more mental effort to maintain the same performance. The second largest pupil appeared in condition 1, as participants were more nervous and excited with heavier cognitive load and arousal. In condition 2, where the temperature and humidity were suitable, a moderate arousal level and pupil diameter were observed. The smallest pupil was found in condition 3, which suggests that it was difficult to keep sufficient arousal in a sleepy state at high temperature and humidity.\par
The result of the heating map revealed that the visual search area in condition 1 was relatively wide, and operators needed to receive more information for adequate situation awareness and understanding, which corresponded to the search pattern under a nervous mode. In condition 2, the search pattern was more precise and operators browsed less in the areas outside the key information which ensured more efficient understanding and better task performance. In condition 3, the participants paid more attention to the confusing information and the areas outside the crucial information, which was consistent with the search pattern under a drowsy state. In condition 4, the fixation area was more chaotic because it was difficult to concentrate in the extremely high temperature and humidity environment. \par
Meanwhile, under the same condition, it was evident that the attention area outside the AOI in the major task 2 is considerably larger than that of the major task 1. This indicated a rapid increase in cognitive load and a decline in performance within a brief period of time (approximately 5 minutes). 

\subsection{The modified maximal adaptability model}
In the range of common temperature and humid conditions, the maximal adaptability model \citep{4hancock2003} has been extensively acknowledged to describe the relationship between performance and environment. Up to now, only a few studies have investigated the performance in the extreme WBGT environment \citep{2Lov1967, 4hancock2003}. However, the model has not been extended to the extreme WBGT environment. We carried out the experiment over 35℃ WBGT (36.81 ℃ WBGT, condition 4) which was close to the human thermal limit within a safe workspace. A temporary simple improved but not sustainable performance was observed, which we attribute to increased cooperation among PFC. Under extreme heat stress, humans respond faster in simple reaction tasks due to stress response, however, load accumulation leads to a rapid decline in complex cognitive functions. The rapid accumulation of workload, both in physical load shown by the highest physical demand score in the NASA-TLX scale and the mental load shown by the biggest pupil from the eye tracking, limited further performance improvement in simple tasks. Additionally, another limitation was the impaired concentration supported by the evidence from the chaotic eye movements and intense ECG activities.

\begin{figure}[H]
    \centering
    \includegraphics[width=0.8\linewidth]{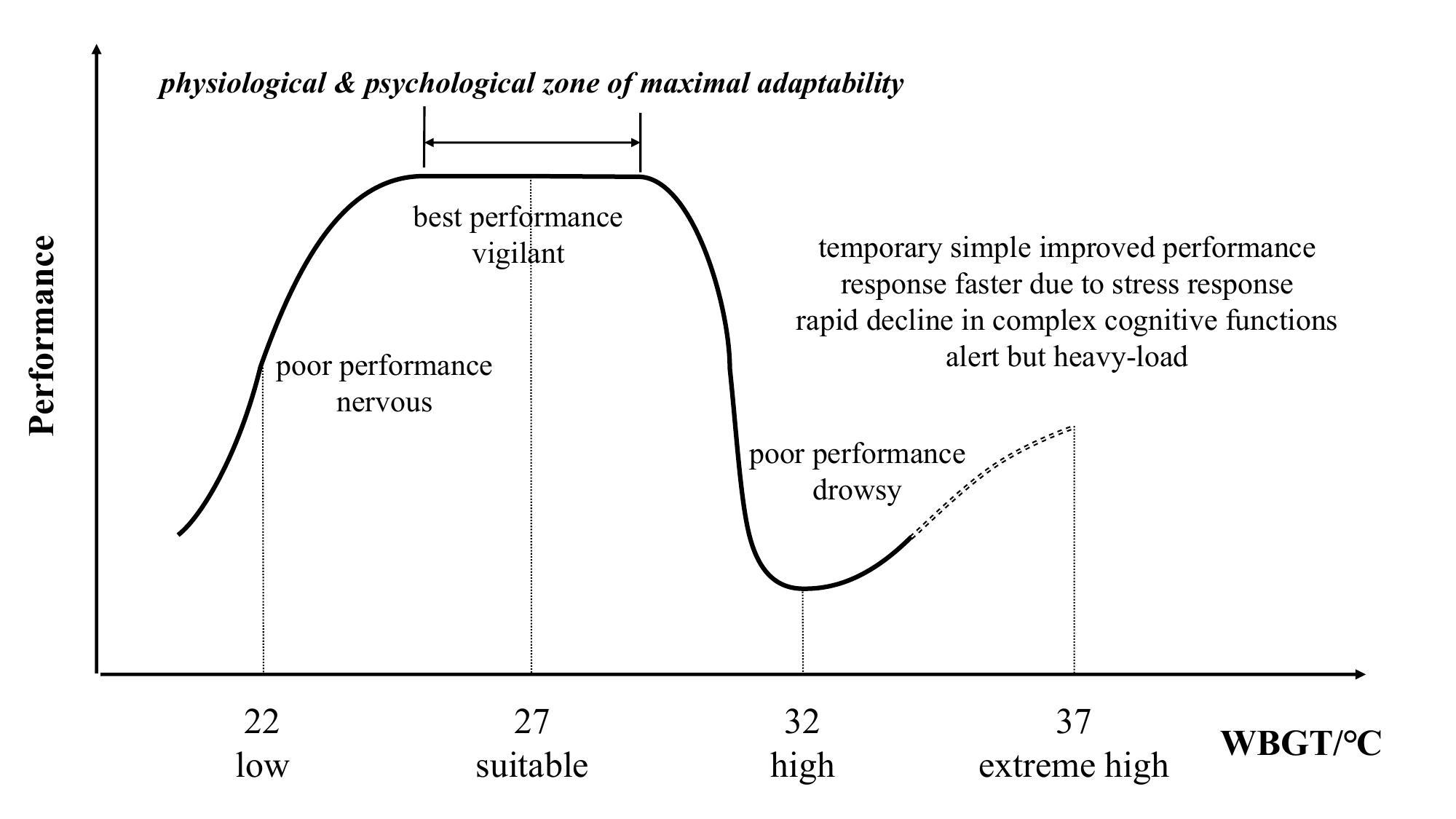}
    \caption{The modified maximal adaptability model extended to the extremely high temperature and humid environment with a temporary improvement in simple reaction tasks but a rapid decline in advanced cognitive functions.}
    \label{fig:modifed model}
\end{figure}

%% file: files/06conclusion.tex
\section{Conclusion}
From a neuroergonomics perspective, this study explored the performance in 21.26 ℃, 26.88 ℃, 31.92 ℃, 36.81 ℃ WBGT and its underlying physiological-psychological mechanism based on ECG, fNIRS, and eye tracking. We verified and extended the maximal adaptability model to extremely high temperature and humid conditions over 35℃ WBGT. Performance under 36.81℃ WBGT could be described and explained by the maximal adaptability model with an extended inverse U shape. Unexpectedly, a temporary simple performance improvement was observed in the hottest and most humid environment. At low temperature and humidity (21.26 ℃ WBGT, condition 1), higher temporal demand in NASA-TLX and imprecise visual search patterns in the heat map showed that bad performance and slower response were caused by nervous mode. In a suitable environment at 26.88 ℃ WBGT (condition 2), enhanced performance with improved working memory and the lowest physical load was observed, which was driven by effective PFC activation and a more accurate search strategy (supported by heat map and greater saccade angle). With growing temperature and humidity, decreasing performance with more mistakes was related to worse understanding and drowsiness at 31.92℃ WBGT (condition 3). The evidence from more activated PFC channels associated with compensatory mechanisms was provided by GLM. More fixation numbers and duration from eye tracking suggested worse information search and processing efficiency. The smallest pupil revealed the lowest arousal and the heat map showed more concentration on confusing information. In the last most extreme environment (36.81 ℃ WBGT, condition 4), there was a temporary simple improvement in simple reaction tasks that could be attributed to more brain connectivity, indicating more cooperation among PFC. The shorter fixation duration and the largest pupil demonstrated an improvement in information processing and arousal. As a result, participants respond faster in simple reaction tasks due to stress response. However, the simple improvement could not be sustained for a long time, and complex cognitive functions (i.e. situation awareness, communication, working memory) were impaired extremely, because of the difficulty in concentration with chaotic visual search heating map and increased vertical eye activity. Evidence from ECG also showed heavier load, poorer regulation ability, and increasing stress, as reflected by faster HR, lower HRV (RMSSD, SDSD, SDNN, PNN50), higher LF/HF, and lower vagal in the extreme environment.  \par
This work provides a more detailed analysis of the process of performance and typical cognitive functions and executive functions under increasing heat stress. This can help to improve personnel management and emergency plans under hazards. In addition, the physiological-psychological mechanism underlying performance and typical functions were explored, which helped understand the interaction between operators' performance and workplace conditions under extreme heat strain, and supported that features extracted by specialized neuroscience methods could be potential biomarkers of cognitive performance under hot-humid exposures. \par
%which is helpful to understand the decision-making mechanism of individuals in an emergency and improve the agent model more precisely in human reliability analysis. \par
This work still has several limitations. Although we extend the maximal adaptability model to extreme heat conditions, four conditions are relatively insufficient to outline an accurate model. Moreover, our work supports a qualitative model which is inconvenient for monitoring the abnormal status of operators during production. In the future, we aim to develop a performance evaluation model based on psychological data and deep-leaning networks to provide a real-time warning system for operators' abnormal states.

%% file: files/07acknowledge.tex
\section{acknowledgment}
This study is supported by the State Key Laboratory of Nuclear Power Safety Technology and Equipment, China Nuclear Power Engineering Co., Ltd. (No.K-A2021.402), and the National Natural Science Foundation of China (No. 72374118, 72304165, 72204136).

%% file: main.bbl
\begin{thebibliography}{52}
\expandafter\ifx\csname natexlab\endcsname\relax\def\natexlab#1{#1}\fi
\providecommand{\url}[1]{\texttt{#1}}
\providecommand{\href}[2]{#2}
\providecommand{\path}[1]{#1}
\providecommand{\DOIprefix}{doi:}
\providecommand{\ArXivprefix}{arXiv:}
\providecommand{\URLprefix}{URL: }
\providecommand{\Pubmedprefix}{pmid:}
\providecommand{\doi}[1]{\href{http://dx.doi.org/#1}{\path{#1}}}
\providecommand{\Pubmed}[1]{\href{pmid:#1}{\path{#1}}}
\providecommand{\bibinfo}[2]{#2}
\ifx\xfnm\relax \def\xfnm[#1]{\unskip,\space#1}\fi
%Type = Article
\bibitem[{thr(2004)}]{threemile2004human}
, \bibinfo{year}{2004}.
\newblock \bibinfo{title}{Human reliability data, human error and accident
  models---illustration through the three mile island accident analysis}.
\newblock \bibinfo{journal}{Reliability Engineering \& System Safety}
  \bibinfo{volume}{83}, \bibinfo{pages}{153--167}.
%Type = Article
\bibitem[{Abbasi et~al.(2020)Abbasi, Motamedzade, Aliabadi, Golmohammadi and
  Tapak}]{64ABBASI2020103189}
\bibinfo{author}{Abbasi, A.M.}, \bibinfo{author}{Motamedzade, M.},
  \bibinfo{author}{Aliabadi, M.}, \bibinfo{author}{Golmohammadi, R.},
  \bibinfo{author}{Tapak, L.}, \bibinfo{year}{2020}.
\newblock \bibinfo{title}{Combined effects of noise and air temperature on
  human neurophysiological responses in a simulated indoor environment}.
\newblock \bibinfo{journal}{Applied Ergonomics} \bibinfo{volume}{88},
  \bibinfo{pages}{103189}.
\newblock \URLprefix
  \url{https://www.sciencedirect.com/science/article/pii/S0003687018303703},
  \DOIprefix\doi{https://doi.org/10.1016/j.apergo.2020.103189}.
%Type = Article
\bibitem[{Ayaz et~al.(2012)Ayaz, Shewokis, Bunce, Izzetoglu, Willems and
  Onaral}]{40ayaz2012optical}
\bibinfo{author}{Ayaz, H.}, \bibinfo{author}{Shewokis, P.A.},
  \bibinfo{author}{Bunce, S.}, \bibinfo{author}{Izzetoglu, K.},
  \bibinfo{author}{Willems, B.}, \bibinfo{author}{Onaral, B.},
  \bibinfo{year}{2012}.
\newblock \bibinfo{title}{Optical brain monitoring for operator training and
  mental workload assessment}.
\newblock \bibinfo{journal}{NeuroImage} \bibinfo{volume}{59},
  \bibinfo{pages}{36--47}.
\newblock \URLprefix
  \url{https://www.sciencedirect.com/science/article/pii/S1053811911006410},
  \DOIprefix\doi{https://doi.org/10.1016/j.neuroimage.2011.06.023}.
  \bibinfo{note}{neuroergonomics: The human brain in action and at work}.
%Type = Article
\bibitem[{Batmaz and Ozturk(2007)}]{31batmaz2008using}
\bibinfo{author}{Batmaz, I.}, \bibinfo{author}{Ozturk, M.},
  \bibinfo{year}{2007}.
\newblock \bibinfo{title}{Using {Pupil} {Diameter} {Changes} for {Measuring}
  {Mental} {Workload} under {Mental} {Processing}}.
\newblock \bibinfo{journal}{Journal of Applied Sciences} \bibinfo{volume}{8},
  \bibinfo{pages}{68--76}.
\newblock \URLprefix
  \url{https://www.scialert.net/abstract/?doi=jas.2008.68.76},
  \DOIprefix\doi{10.3923/jas.2008.68.76}.
%Type = Article
\bibitem[{Causse et~al.(2017)Causse, Chua, Peysakhovich, Del~Campo and
  Matton}]{38causse2017mental}
\bibinfo{author}{Causse, M.}, \bibinfo{author}{Chua, Z.},
  \bibinfo{author}{Peysakhovich, V.}, \bibinfo{author}{Del~Campo, N.},
  \bibinfo{author}{Matton, N.}, \bibinfo{year}{2017}.
\newblock \bibinfo{title}{Mental workload and neural efficiency quantified in
  the prefrontal cortex using {fNIRS}}.
\newblock \bibinfo{journal}{Scientific Reports} \bibinfo{volume}{7},
  \bibinfo{pages}{5222}.
\newblock \URLprefix \url{https://www.nature.com/articles/s41598-017-05378-x},
  \DOIprefix\doi{10.1038/s41598-017-05378-x}.
%Type = Article
\bibitem[{Chang et~al.(2017)Chang, Bernard and Logan}]{61riskCHANG2017150}
\bibinfo{author}{Chang, C.H.}, \bibinfo{author}{Bernard, T.E.},
  \bibinfo{author}{Logan, J.}, \bibinfo{year}{2017}.
\newblock \bibinfo{title}{Effects of heat stress on risk perceptions and risk
  taking}.
\newblock \bibinfo{journal}{Applied Ergonomics} \bibinfo{volume}{62},
  \bibinfo{pages}{150--157}.
\newblock \URLprefix
  \url{https://www.sciencedirect.com/science/article/pii/S0003687017300546},
  \DOIprefix\doi{https://doi.org/10.1016/j.apergo.2017.02.018}.
%Type = Article
\bibitem[{Chen et~al.(2020)Chen, Tao and Liu}]{65CHEN2020107372}
\bibinfo{author}{Chen, Y.}, \bibinfo{author}{Tao, M.}, \bibinfo{author}{Liu,
  W.}, \bibinfo{year}{2020}.
\newblock \bibinfo{title}{High temperature impairs cognitive performance during
  a moderate intensity activity}.
\newblock \bibinfo{journal}{Building and Environment} \bibinfo{volume}{186},
  \bibinfo{pages}{107372}.
\newblock \URLprefix
  \url{https://www.sciencedirect.com/science/article/pii/S0360132320307411},
  \DOIprefix\doi{https://doi.org/10.1016/j.buildenv.2020.107372}.
%Type = Article
\bibitem[{Chen et~al.(2023)Chen, Wang, Tian and Liu}]{heartU_CHEN2023109801}
\bibinfo{author}{Chen, Y.}, \bibinfo{author}{Wang, Z.}, \bibinfo{author}{Tian,
  X.}, \bibinfo{author}{Liu, W.}, \bibinfo{year}{2023}.
\newblock \bibinfo{title}{Evaluation of cognitive performance in high
  temperature with heart rate: A pilot study}.
\newblock \bibinfo{journal}{Building and Environment} \bibinfo{volume}{228},
  \bibinfo{pages}{109801}.
\newblock \URLprefix
  \url{https://www.sciencedirect.com/science/article/pii/S0360132322010319},
  \DOIprefix\doi{https://doi.org/10.1016/j.buildenv.2022.109801}.
%Type = Article
\bibitem[{Cole et~al.(2012)Cole, Yarkoni, Repov{\v{s}}, Anticevic and
  Braver}]{44cole2012global}
\bibinfo{author}{Cole, M.W.}, \bibinfo{author}{Yarkoni, T.},
  \bibinfo{author}{Repov{\v{s}}, G.}, \bibinfo{author}{Anticevic, A.},
  \bibinfo{author}{Braver, T.S.}, \bibinfo{year}{2012}.
\newblock \bibinfo{title}{Global connectivity of prefrontal cortex predicts
  cognitive control and intelligence}.
\newblock \bibinfo{journal}{Journal of Neuroscience} \bibinfo{volume}{32},
  \bibinfo{pages}{8988--8999}.
%Type = Article
\bibitem[{De~Rivecourt et~al.(2008)De~Rivecourt, Kuperus, Post and
  Mulder}]{24de2008cardiovascular}
\bibinfo{author}{De~Rivecourt, M.}, \bibinfo{author}{Kuperus, M.N.},
  \bibinfo{author}{Post, W.J.}, \bibinfo{author}{Mulder, L.J.},
  \bibinfo{year}{2008}.
\newblock \bibinfo{title}{Cardiovascular and eye activity measures as indices
  for momentary changes in mental effort during simulated flight}.
\newblock \bibinfo{journal}{Ergonomics} \bibinfo{volume}{51},
  \bibinfo{pages}{1295--1319}.
\newblock \URLprefix
  \url{https://www.tandfonline.com/doi/full/10.1080/00140130802120267},
  \DOIprefix\doi{10.1080/00140130802120267}.
%Type = Article
\bibitem[{Delaney and Brodie(2000)}]{25delaney2000effects}
\bibinfo{author}{Delaney, J.P.A.}, \bibinfo{author}{Brodie, D.A.},
  \bibinfo{year}{2000}.
\newblock \bibinfo{title}{Effects of {Short}-{Term} {Psychological} {Stress} on
  the {Time} and {Frequency} {Domains} of {Heart}-{Rate} {Variability}}.
\newblock \bibinfo{journal}{Perceptual and Motor Skills} \bibinfo{volume}{91},
  \bibinfo{pages}{515--524}.
\newblock \URLprefix
  \url{http://journals.sagepub.com/doi/10.2466/pms.2000.91.2.515},
  \DOIprefix\doi{10.2466/pms.2000.91.2.515}.
%Type = Inproceedings
\bibitem[{Endsley(1988)}]{SAGATendsley1988situation}
\bibinfo{author}{Endsley, M.}, \bibinfo{year}{1988}.
\newblock \bibinfo{title}{Situation awareness global assessment technique
  (sagat)}, in: \bibinfo{booktitle}{Proceedings of the IEEE 1988 National
  Aerospace and Electronics Conference}, pp. \bibinfo{pages}{789--795 vol.3}.
\newblock \DOIprefix\doi{10.1109/NAECON.1988.195097}.
%Type = Article
\bibitem[{Gaoua et~al.(2012)Gaoua, Grantham, Racinais and {El
  Massioui}}]{57gaoua2012sensory}
\bibinfo{author}{Gaoua, N.}, \bibinfo{author}{Grantham, J.},
  \bibinfo{author}{Racinais, S.}, \bibinfo{author}{{El Massioui}, F.},
  \bibinfo{year}{2012}.
\newblock \bibinfo{title}{Sensory displeasure reduces complex cognitive
  performance in the heat}.
\newblock \bibinfo{journal}{Journal of Environmental Psychology}
  \bibinfo{volume}{32}, \bibinfo{pages}{158--163}.
\newblock \URLprefix
  \url{https://www.sciencedirect.com/science/article/pii/S0272494412000035},
  \DOIprefix\doi{https://doi.org/10.1016/j.jenvp.2012.01.002}.
%Type = Article
\bibitem[{Gaoua et~al.(2011)Gaoua, Racinais, Grantham and
  El~Massioui}]{56gaoua2011alterations}
\bibinfo{author}{Gaoua, N.}, \bibinfo{author}{Racinais, S.},
  \bibinfo{author}{Grantham, J.}, \bibinfo{author}{El~Massioui, F.},
  \bibinfo{year}{2011}.
\newblock \bibinfo{title}{Alterations in cognitive performance during passive
  hyperthermia are task dependent}.
\newblock \bibinfo{journal}{International Journal of Hyperthermia}
  \bibinfo{volume}{27}, \bibinfo{pages}{1--9}.
\newblock \URLprefix
  \url{http://www.tandfonline.com/doi/full/10.3109/02656736.2010.516305},
  \DOIprefix\doi{10.3109/02656736.2010.516305}.
%Type = Article
\bibitem[{Guasti et~al.(2005)Guasti, Simoni, Mainardi, Crespi, Cimpanelli,
  Klersy, Gaudio, Grandi, Cerutti and Venco}]{27guasti2005global}
\bibinfo{author}{Guasti, L.}, \bibinfo{author}{Simoni, C.},
  \bibinfo{author}{Mainardi, L.}, \bibinfo{author}{Crespi, C.},
  \bibinfo{author}{Cimpanelli, M.}, \bibinfo{author}{Klersy, C.},
  \bibinfo{author}{Gaudio, G.}, \bibinfo{author}{Grandi, A.M.},
  \bibinfo{author}{Cerutti, S.}, \bibinfo{author}{Venco, A.},
  \bibinfo{year}{2005}.
\newblock \bibinfo{title}{Global link between heart rate and blood pressure
  oscillations at rest and during mental arousal in normotensive and
  hypertensive subjects}.
\newblock \bibinfo{journal}{Autonomic Neuroscience} \bibinfo{volume}{120},
  \bibinfo{pages}{80--87}.
\newblock \URLprefix
  \url{https://www.sciencedirect.com/science/article/pii/S1566070205000342},
  \DOIprefix\doi{https://doi.org/10.1016/j.autneu.2005.02.008}.
%Type = Article
\bibitem[{Hancock(1989)}]{maximodel_hancock_dynamic_1989}
\bibinfo{author}{Hancock, P.A.}, \bibinfo{year}{1989}.
\newblock \bibinfo{title}{A {Dynamic} {Model} of {Stress} and {Sustained}
  {Attention}}.
\newblock \bibinfo{journal}{Human Factors: The Journal of the Human Factors and
  Ergonomics Society} \bibinfo{volume}{31}, \bibinfo{pages}{519--537}.
\newblock \URLprefix
  \url{http://journals.sagepub.com/doi/10.1177/001872088903100503},
  \DOIprefix\doi{10.1177/001872088903100503}.
%Type = Article
\bibitem[{Hancock and Vasmatzidis(2003)}]{4hancock2003}
\bibinfo{author}{Hancock, P.A.}, \bibinfo{author}{Vasmatzidis, I.},
  \bibinfo{year}{2003}.
\newblock \bibinfo{title}{Effects of heat stress on cognitive performance: the
  current state of knowledge}.
\newblock \bibinfo{journal}{International Journal of Hyperthermia}
  \bibinfo{volume}{19}, \bibinfo{pages}{355--372}.
\newblock \URLprefix
  \url{https://www.tandfonline.com/doi/full/10.1080/0265673021000054630},
  \DOIprefix\doi{10.1080/0265673021000054630}.
%Type = Article
\bibitem[{Hou et~al.(2021)Hou, Zhang, Zhao, Duan, Gong, Li and Zhu}]{nirskit}
\bibinfo{author}{Hou, X.}, \bibinfo{author}{Zhang, Z.}, \bibinfo{author}{Zhao,
  C.}, \bibinfo{author}{Duan, L.}, \bibinfo{author}{Gong, Y.},
  \bibinfo{author}{Li, Z.}, \bibinfo{author}{Zhu, C.}, \bibinfo{year}{2021}.
\newblock \bibinfo{title}{{NIRS}-{KIT}: a {MATLAB} toolbox for both
  resting-state and task {fNIRS} data analysis}.
\newblock \bibinfo{journal}{Neurophotonics} \bibinfo{volume}{8}.
\newblock \URLprefix
  \url{https://www.spiedigitallibrary.org/journals/neurophotonics/volume-8/issue-01/010802/NIRS-KIT--a-MATLAB-toolbox-for-both-resting-state/10.1117/1.NPh.8.1.010802.full},
  \DOIprefix\doi{10.1117/1.NPh.8.1.010802}.
%Type = Article
\bibitem[{Hwang et~al.(2008)Hwang, Yau, Lin, Chen, Huang, Yenn and
  Hsu}]{23hwang2008predicting}
\bibinfo{author}{Hwang, S.L.}, \bibinfo{author}{Yau, Y.J.},
  \bibinfo{author}{Lin, Y.T.}, \bibinfo{author}{Chen, J.H.},
  \bibinfo{author}{Huang, T.H.}, \bibinfo{author}{Yenn, T.C.},
  \bibinfo{author}{Hsu, C.C.}, \bibinfo{year}{2008}.
\newblock \bibinfo{title}{Predicting work performance in nuclear power plants}.
\newblock \bibinfo{journal}{Safety Science} \bibinfo{volume}{46},
  \bibinfo{pages}{1115--1124}.
\newblock \URLprefix
  \url{https://www.sciencedirect.com/science/article/pii/S0925753507000872},
  \DOIprefix\doi{https://doi.org/10.1016/j.ssci.2007.06.005}.
%Type = Article
\bibitem[{Jiang et~al.(2013)Jiang, Yang, Liu, Li, Li, Li, Qian, Zhao, Zhou and
  Sun}]{fmri4_jiang_hyperthermia_2013}
\bibinfo{author}{Jiang, Q.}, \bibinfo{author}{Yang, X.}, \bibinfo{author}{Liu,
  K.}, \bibinfo{author}{Li, B.}, \bibinfo{author}{Li, L.}, \bibinfo{author}{Li,
  M.}, \bibinfo{author}{Qian, S.}, \bibinfo{author}{Zhao, L.},
  \bibinfo{author}{Zhou, Z.}, \bibinfo{author}{Sun, G.}, \bibinfo{year}{2013}.
\newblock \bibinfo{title}{Hyperthermia impaired human visual short-term memory:
  {An} {fMRI} study}.
\newblock \bibinfo{journal}{International Journal of Hyperthermia}
  \bibinfo{volume}{29}, \bibinfo{pages}{219--224}.
\newblock \URLprefix
  \url{http://www.tandfonline.com/doi/full/10.3109/02656736.2013.786141},
  \DOIprefix\doi{10.3109/02656736.2013.786141}.
%Type = Article
\bibitem[{Kim et~al.(2020)Kim, Hong, Kim and Yeom}]{kim2020psychophysiological}
\bibinfo{author}{Kim, H.}, \bibinfo{author}{Hong, T.}, \bibinfo{author}{Kim,
  J.}, \bibinfo{author}{Yeom, S.}, \bibinfo{year}{2020}.
\newblock \bibinfo{title}{A psychophysiological effect of indoor thermal
  condition on college students’ learning performance through eeg
  measurement}.
\newblock \bibinfo{journal}{Building and Environment} \bibinfo{volume}{184},
  \bibinfo{pages}{107223}.
\newblock \URLprefix
  \url{https://www.sciencedirect.com/science/article/pii/S0360132320305941},
  \DOIprefix\doi{https://doi.org/10.1016/j.buildenv.2020.107223}.
%Type = Article
\bibitem[{Lan et~al.(2010)Lan, Lian and Pan}]{63LAN201029}
\bibinfo{author}{Lan, L.}, \bibinfo{author}{Lian, Z.}, \bibinfo{author}{Pan,
  L.}, \bibinfo{year}{2010}.
\newblock \bibinfo{title}{The effects of air temperature on office workers’
  well-being, workload and productivity-evaluated with subjective ratings}.
\newblock \bibinfo{journal}{Applied Ergonomics} \bibinfo{volume}{42},
  \bibinfo{pages}{29--36}.
\newblock \URLprefix
  \url{https://www.sciencedirect.com/science/article/pii/S000368701000058X},
  \DOIprefix\doi{https://doi.org/10.1016/j.apergo.2010.04.003}.
%Type = Article
\bibitem[{Liu et~al.(2021)Liu, Zhang, Sun, Gao, Jing and
  Ye}]{68pupil_LIU2021101458}
\bibinfo{author}{Liu, C.}, \bibinfo{author}{Zhang, Y.}, \bibinfo{author}{Sun,
  L.}, \bibinfo{author}{Gao, W.}, \bibinfo{author}{Jing, X.},
  \bibinfo{author}{Ye, W.}, \bibinfo{year}{2021}.
\newblock \bibinfo{title}{Influence of indoor air temperature and relative
  humidity on learning performance of undergraduates}.
\newblock \bibinfo{journal}{Case Studies in Thermal Engineering}
  \bibinfo{volume}{28}, \bibinfo{pages}{101458}.
\newblock \URLprefix
  \url{https://www.sciencedirect.com/science/article/pii/S2214157X21006213},
  \DOIprefix\doi{https://doi.org/10.1016/j.csite.2021.101458}.
%Type = Article
\bibitem[{Liu et~al.(2013)Liu, Sun, Li, Jiang, Yang, Li, Li, Qian, Zhao, Zhou,
  {von Deneen} and Liu}]{fmri3_LIU2013220}
\bibinfo{author}{Liu, K.}, \bibinfo{author}{Sun, G.}, \bibinfo{author}{Li, B.},
  \bibinfo{author}{Jiang, Q.}, \bibinfo{author}{Yang, X.}, \bibinfo{author}{Li,
  M.}, \bibinfo{author}{Li, L.}, \bibinfo{author}{Qian, S.},
  \bibinfo{author}{Zhao, L.}, \bibinfo{author}{Zhou, Z.}, \bibinfo{author}{{von
  Deneen}, K.M.}, \bibinfo{author}{Liu, Y.}, \bibinfo{year}{2013}.
\newblock \bibinfo{title}{The impact of passive hyperthermia on human attention
  networks: An fmri study}.
\newblock \bibinfo{journal}{Behavioural Brain Research} \bibinfo{volume}{243},
  \bibinfo{pages}{220--230}.
\newblock \URLprefix
  \url{https://www.sciencedirect.com/science/article/pii/S0166432813000235},
  \DOIprefix\doi{https://doi.org/10.1016/j.bbr.2013.01.013}.
%Type = Article
\bibitem[{Liu et~al.(2022)Liu, Tian and Tao}]{67expU_LIU2022108431}
\bibinfo{author}{Liu, W.}, \bibinfo{author}{Tian, X.}, \bibinfo{author}{Tao,
  M.}, \bibinfo{year}{2022}.
\newblock \bibinfo{title}{A model to quantify the relation between cognitive
  performance and thermal responses in high temperature at a moderate activity
  level}.
\newblock \bibinfo{journal}{Building and Environment} \bibinfo{volume}{207},
  \bibinfo{pages}{108431}.
\newblock \URLprefix
  \url{https://www.sciencedirect.com/science/article/pii/S0360132321008283},
  \DOIprefix\doi{https://doi.org/10.1016/j.buildenv.2021.108431}.
%Type = Article
\bibitem[{Lovingood et~al.(1967)Lovingood, Blyth, Peacock and
  Lindsay}]{2Lov1967}
\bibinfo{author}{Lovingood, B.W.}, \bibinfo{author}{Blyth, C.S.},
  \bibinfo{author}{Peacock, W.H.}, \bibinfo{author}{Lindsay, R.B.},
  \bibinfo{year}{1967}.
\newblock \bibinfo{title}{Effects of d-{Amphetamine} {Sulfate}, {Caffeine}, and
  {High} {Temperature} on {Human} {Performance}}.
\newblock \bibinfo{journal}{Research Quarterly. American Association for
  Health, Physical Education and Recreation} \bibinfo{volume}{38},
  \bibinfo{pages}{64--71}.
\newblock \URLprefix
  \url{https://www.tandfonline.com/doi/full/10.1080/10671188.1967.10614804},
  \DOIprefix\doi{10.1080/10671188.1967.10614804}.
%Type = Article
\bibitem[{Malmo and Malmo(2000)}]{1malmo2000electromyographic}
\bibinfo{author}{Malmo, R.B.}, \bibinfo{author}{Malmo, H.P.},
  \bibinfo{year}{2000}.
\newblock \bibinfo{title}{On electromyographic (emg) gradients and
  movement-related brain activity: significance for motor control, cognitive
  functions, and certain psychopathologies}.
\newblock \bibinfo{journal}{International Journal of Psychophysiology}
  \bibinfo{volume}{38}, \bibinfo{pages}{143--207}.
\newblock \URLprefix
  \url{https://www.sciencedirect.com/science/article/pii/S0167876000001136},
  \DOIprefix\doi{https://doi.org/10.1016/S0167-8760(00)00113-6}.
%Type = Article
\bibitem[{Martelli et~al.(2014)Martelli, Silvani, McAllen, May and
  Ramchandra}]{26martelli2014low}
\bibinfo{author}{Martelli, D.}, \bibinfo{author}{Silvani, A.},
  \bibinfo{author}{McAllen, R.M.}, \bibinfo{author}{May, C.N.},
  \bibinfo{author}{Ramchandra, R.}, \bibinfo{year}{2014}.
\newblock \bibinfo{title}{The low frequency power of heart rate variability is
  neither a measure of cardiac sympathetic tone nor of baroreflex sensitivity}.
\newblock \bibinfo{journal}{American Journal of Physiology-Heart and
  Circulatory Physiology} \bibinfo{volume}{307}, \bibinfo{pages}{H1005--H1012}.
\newblock \URLprefix
  \url{https://www.physiology.org/doi/10.1152/ajpheart.00361.2014},
  \DOIprefix\doi{10.1152/ajpheart.00361.2014}.
%Type = Article
\bibitem[{Matsuda and Hiraki(2006)}]{41matsuda2006sustained}
\bibinfo{author}{Matsuda, G.}, \bibinfo{author}{Hiraki, K.},
  \bibinfo{year}{2006}.
\newblock \bibinfo{title}{Sustained decrease in oxygenated hemoglobin during
  video games in the dorsal prefrontal cortex: A nirs study of children}.
\newblock \bibinfo{journal}{NeuroImage} \bibinfo{volume}{29},
  \bibinfo{pages}{706--711}.
\newblock \URLprefix
  \url{https://www.sciencedirect.com/science/article/pii/S1053811905006233},
  \DOIprefix\doi{https://doi.org/10.1016/j.neuroimage.2005.08.019}.
%Type = Article
\bibitem[{Nakata et~al.(2021)Nakata, Kakigi and
  Shibasaki}]{66nakata_effects_2021}
\bibinfo{author}{Nakata, H.}, \bibinfo{author}{Kakigi, R.},
  \bibinfo{author}{Shibasaki, M.}, \bibinfo{year}{2021}.
\newblock \bibinfo{title}{Effects of passive heat stress and recovery on human
  cognitive function: {An} {ERP} study}.
\newblock \bibinfo{journal}{PLOS ONE} \bibinfo{volume}{16},
  \bibinfo{pages}{e0254769}.
\newblock \URLprefix \url{https://dx.plos.org/10.1371/journal.pone.0254769},
  \DOIprefix\doi{10.1371/journal.pone.0254769}.
%Type = Inproceedings
\bibitem[{Nakayama et~al.(2002)Nakayama, Takahashi and
  Shimizu}]{36nakayama2002act}
\bibinfo{author}{Nakayama, M.}, \bibinfo{author}{Takahashi, K.},
  \bibinfo{author}{Shimizu, Y.}, \bibinfo{year}{2002}.
\newblock \bibinfo{title}{The act of task difficulty and eye-movement frequency
  for the '{Oculo}-motor indices'}, in: \bibinfo{booktitle}{Proceedings of the
  symposium on {Eye} tracking research \& applications - {ETRA} '02},
  \bibinfo{publisher}{ACM Press}, \bibinfo{address}{New Orleans, Louisiana}.
  p.~\bibinfo{pages}{37}.
\newblock \URLprefix
  \url{http://portal.acm.org/citation.cfm?doid=507072.507080},
  \DOIprefix\doi{10.1145/507072.507080}.
%Type = Article
\bibitem[{Neubauer and Fink(2009)}]{42neubauer2009intelligence}
\bibinfo{author}{Neubauer, A.C.}, \bibinfo{author}{Fink, A.},
  \bibinfo{year}{2009}.
\newblock \bibinfo{title}{Intelligence and neural efficiency}.
\newblock \bibinfo{journal}{Neuroscience \& Biobehavioral Reviews}
  \bibinfo{volume}{33}, \bibinfo{pages}{1004--1023}.
\newblock \URLprefix
  \url{https://www.sciencedirect.com/science/article/pii/S0149763409000591},
  \DOIprefix\doi{https://doi.org/10.1016/j.neubiorev.2009.04.001}.
%Type = Article
\bibitem[{O’Hara et~al.(2008)O’Hara, Higgins, Brown, Fink, Persensky,
  Lewis, Kramer, Szabo and Boggi}]{cogmodel1ohara2008human}
\bibinfo{author}{O’Hara, J.}, \bibinfo{author}{Higgins, J.},
  \bibinfo{author}{Brown, W.}, \bibinfo{author}{Fink, R.},
  \bibinfo{author}{Persensky, J.}, \bibinfo{author}{Lewis, P.},
  \bibinfo{author}{Kramer, J.}, \bibinfo{author}{Szabo, A.},
  \bibinfo{author}{Boggi, M.}, \bibinfo{year}{2008}.
\newblock \bibinfo{title}{Human factors considerations with respect to emerging
  technology in nuclear power plants}.
\newblock \bibinfo{journal}{US Nuclear Regulatory Commission, Washington, DC,
  USA} .
%Type = Article
\bibitem[{Pakarinen et~al.(2018)Pakarinen, Korpela, Torniainen, Laarni and
  Karvonen}]{28pakarinen2018cardiac}
\bibinfo{author}{Pakarinen, S.}, \bibinfo{author}{Korpela, J.},
  \bibinfo{author}{Torniainen, J.}, \bibinfo{author}{Laarni, J.},
  \bibinfo{author}{Karvonen, H.}, \bibinfo{year}{2018}.
\newblock \bibinfo{title}{Cardiac measures of nuclear power plant operator
  stress during simulated incident and accident scenarios}.
\newblock \bibinfo{journal}{Psychophysiology} \bibinfo{volume}{55},
  \bibinfo{pages}{e13071}.
\newblock \URLprefix
  \url{https://onlinelibrary.wiley.com/doi/10.1111/psyp.13071},
  \DOIprefix\doi{10.1111/psyp.13071}.
%Type = Article
\bibitem[{Pan et~al.(2019)Pan, Borragán and Peigneux}]{39pan2019applications}
\bibinfo{author}{Pan, Y.}, \bibinfo{author}{Borragán, G.},
  \bibinfo{author}{Peigneux, P.}, \bibinfo{year}{2019}.
\newblock \bibinfo{title}{Applications of {Functional} {Near}-{Infrared}
  {Spectroscopy} in {Fatigue}, {Sleep} {Deprivation}, and {Social}
  {Cognition}}.
\newblock \bibinfo{journal}{Brain Topography} \bibinfo{volume}{32},
  \bibinfo{pages}{998--1012}.
\newblock \URLprefix \url{http://link.springer.com/10.1007/s10548-019-00740-w},
  \DOIprefix\doi{10.1007/s10548-019-00740-w}.
%Type = Article
\bibitem[{Pilcher et~al.(2002)Pilcher, Nadler and Busch}]{5pilcher2002effects}
\bibinfo{author}{Pilcher, J.J.}, \bibinfo{author}{Nadler, E.},
  \bibinfo{author}{Busch, C.}, \bibinfo{year}{2002}.
\newblock \bibinfo{title}{Effects of hot and cold temperature exposure on
  performance: a meta-analytic review}.
\newblock \bibinfo{journal}{Ergonomics} \bibinfo{volume}{45},
  \bibinfo{pages}{682--698}.
\newblock \URLprefix
  \url{https://www.tandfonline.com/doi/full/10.1080/00140130210158419},
  \DOIprefix\doi{10.1080/00140130210158419}.
%Type = Article
\bibitem[{Provins(1966)}]{inverU_provins_environmental_1966}
\bibinfo{author}{Provins, K.A.}, \bibinfo{year}{1966}.
\newblock \bibinfo{title}{Environmental heat, body temperature and behaviour:
  {An} hypothesis}.
\newblock \bibinfo{journal}{Australian Journal of Psychology}
  \bibinfo{volume}{18}, \bibinfo{pages}{118--129}.
\newblock \URLprefix
  \url{https://www.tandfonline.com/doi/full/10.1080/00049536608255722},
  \DOIprefix\doi{10.1080/00049536608255722}.
%Type = Article
\bibitem[{Qian et~al.(2020)Qian, Zhang, Yan, Shi, Wang and
  Zhou}]{fmri2_qian_disrupted_2020}
\bibinfo{author}{Qian, S.}, \bibinfo{author}{Zhang, J.}, \bibinfo{author}{Yan,
  S.}, \bibinfo{author}{Shi, Z.}, \bibinfo{author}{Wang, Z.},
  \bibinfo{author}{Zhou, Y.}, \bibinfo{year}{2020}.
\newblock \bibinfo{title}{Disrupted {Anti}-correlation {Between} the {Default}
  and {Dorsal} {Attention} {Networks} {During} {Hyperthermia} {Exposure}: {An}
  {fMRI} {Study}}.
\newblock \bibinfo{journal}{Frontiers in Human Neuroscience}
  \bibinfo{volume}{14}, \bibinfo{pages}{564272}.
\newblock \URLprefix
  \url{https://www.frontiersin.org/articles/10.3389/fnhum.2020.564272/full},
  \DOIprefix\doi{10.3389/fnhum.2020.564272}.
%Type = Article
\bibitem[{Ramsey(1995)}]{3ramsey1995}
\bibinfo{author}{Ramsey, J.D.}, \bibinfo{year}{1995}.
\newblock \bibinfo{title}{Task performance in heat: a review}.
\newblock \bibinfo{journal}{Ergonomics} \bibinfo{volume}{38},
  \bibinfo{pages}{154--165}.
\newblock \URLprefix
  \url{http://www.tandfonline.com/doi/abs/10.1080/00140139508925092},
  \DOIprefix\doi{10.1080/00140139508925092}.
%Type = Article
\bibitem[{Shaffer et~al.(2014)Shaffer, McCraty and Zerr}]{22shaffer2014healthy}
\bibinfo{author}{Shaffer, F.}, \bibinfo{author}{McCraty, R.},
  \bibinfo{author}{Zerr, C.L.}, \bibinfo{year}{2014}.
\newblock \bibinfo{title}{A healthy heart is not a metronome: an integrative
  review of the heart's anatomy and heart rate variability}.
\newblock \bibinfo{journal}{Frontiers in Psychology} \bibinfo{volume}{5}.
\newblock \URLprefix
  \url{http://journal.frontiersin.org/article/10.3389/fpsyg.2014.01040/abstract},
  \DOIprefix\doi{10.3389/fpsyg.2014.01040}.
%Type = Book
\bibitem[{for Standardization(2017)}]{0iso}
\bibinfo{author}{for Standardization, I.O.}, \bibinfo{year}{2017}.
\newblock \bibinfo{title}{Ergonomics of the thermal environment — Assessment
  of heat stress using the WBGT (wet bulb globe temperature) index}.
\newblock \bibinfo{edition}{{ISO 7243:2017}} ed.,
  \bibinfo{publisher}{International Organization for Standardization},
  \bibinfo{address}{Vernier, Geneva, Switzerland}.
\newblock \URLprefix \url{https://www.iso.org/standard/63098.html}.
%Type = Article
\bibitem[{Tan et~al.(2023)Tan, Stephenson, Alhadad, Loh, Soong, Lee and
  Low}]{fmri1_TAN2023}
\bibinfo{author}{Tan, X.R.}, \bibinfo{author}{Stephenson, M.C.},
  \bibinfo{author}{Alhadad, S.B.}, \bibinfo{author}{Loh, K.W.},
  \bibinfo{author}{Soong, T.W.}, \bibinfo{author}{Lee, J.K.},
  \bibinfo{author}{Low, I.C.}, \bibinfo{year}{2023}.
\newblock \bibinfo{title}{Elevated brain temperature under severe heat exposure
  impairs cortical motor activity and executive function}.
\newblock \bibinfo{journal}{Journal of Sport and Health Science} \URLprefix
  \url{https://www.sciencedirect.com/science/article/pii/S2095254623000789},
  \DOIprefix\doi{https://doi.org/10.1016/j.jshs.2023.09.001}.
%Type = Article
\bibitem[{Tian et~al.(2021)Tian, Fang and Liu}]{7tian2021decreased}
\bibinfo{author}{Tian, X.}, \bibinfo{author}{Fang, Z.}, \bibinfo{author}{Liu,
  W.}, \bibinfo{year}{2021}.
\newblock \bibinfo{title}{Decreased humidity improves cognitive performance at
  extreme high indoor temperature}.
\newblock \bibinfo{journal}{Indoor Air} \bibinfo{volume}{31},
  \bibinfo{pages}{608--627}.
\newblock \URLprefix
  \url{https://onlinelibrary.wiley.com/doi/10.1111/ina.12755},
  \DOIprefix\doi{10.1111/ina.12755}.
%Type = Inproceedings
\bibitem[{Tran et~al.(2017)Tran, Yan, Habiyaremye and
  Wei}]{32tran2017predicting}
\bibinfo{author}{Tran, C.C.}, \bibinfo{author}{Yan, S.},
  \bibinfo{author}{Habiyaremye, J.L.}, \bibinfo{author}{Wei, Y.},
  \bibinfo{year}{2017}.
\newblock \bibinfo{title}{Predicting driver’s work performance in driving
  simulator based on physiological indices}, in:
  \bibinfo{booktitle}{Intelligent Human Computer Interaction: 9th International
  Conference, IHCI 2017, Evry, France, December 11-13, 2017, Proceedings 9},
  \bibinfo{organization}{Springer International Publishing}. pp.
  \bibinfo{pages}{150--162}.
%Type = Article
\bibitem[{Van Der~Wel and Van~Steenbergen(2018)}]{30van2018pupil}
\bibinfo{author}{Van Der~Wel, P.}, \bibinfo{author}{Van~Steenbergen, H.},
  \bibinfo{year}{2018}.
\newblock \bibinfo{title}{Pupil dilation as an index of effort in cognitive
  control tasks: {A} review}.
\newblock \bibinfo{journal}{Psychonomic Bulletin \& Review}
  \bibinfo{volume}{25}, \bibinfo{pages}{2005--2015}.
\newblock \URLprefix \url{http://link.springer.com/10.3758/s13423-018-1432-y},
  \DOIprefix\doi{10.3758/s13423-018-1432-y}.
%Type = Article
\bibitem[{Visnovcova et~al.(2016)Visnovcova, Mestanik, Gala, Mestanikova and
  Tonhajzerova}]{52visnovcova2016complexity}
\bibinfo{author}{Visnovcova, Z.}, \bibinfo{author}{Mestanik, M.},
  \bibinfo{author}{Gala, M.}, \bibinfo{author}{Mestanikova, A.},
  \bibinfo{author}{Tonhajzerova, I.}, \bibinfo{year}{2016}.
\newblock \bibinfo{title}{The complexity of electrodermal activity is altered
  in mental cognitive stressors}.
\newblock \bibinfo{journal}{Computers in Biology and Medicine}
  \bibinfo{volume}{79}, \bibinfo{pages}{123--129}.
\newblock \URLprefix
  \url{https://www.sciencedirect.com/science/article/pii/S0010482516302724},
  \DOIprefix\doi{https://doi.org/10.1016/j.compbiomed.2016.10.014}.
%Type = Book
\bibitem[{Whaley(2016)}]{cogmodel2whaley2016cognitive}
\bibinfo{author}{Whaley, A.M.}, \bibinfo{year}{2016}.
\newblock \bibinfo{title}{Cognitive basis for human reliability analysis}.
\newblock \bibinfo{publisher}{US Nuclear Regulatory Commission, Office of
  Nuclear Regulatory Research}.
%Type = Article
\bibitem[{Wu et~al.(2020)Wu, Liu, Jia, Tran and Yan}]{34wu2020using}
\bibinfo{author}{Wu, Y.}, \bibinfo{author}{Liu, Z.}, \bibinfo{author}{Jia, M.},
  \bibinfo{author}{Tran, C.C.}, \bibinfo{author}{Yan, S.},
  \bibinfo{year}{2020}.
\newblock \bibinfo{title}{Using {Artificial} {Neural} {Networks} for
  {Predicting} {Mental} {Workload} in {Nuclear} {Power} {Plants} {Based} on
  {Eye} {Tracking}}.
\newblock \bibinfo{journal}{Nuclear Technology} \bibinfo{volume}{206},
  \bibinfo{pages}{94--106}.
\newblock \URLprefix
  \url{https://www.tandfonline.com/doi/full/10.1080/00295450.2019.1620055},
  \DOIprefix\doi{10.1080/00295450.2019.1620055}.
%Type = Article
\bibitem[{Xu et~al.(2021)Xu, Lu, Vogel-Heuser and Wang}]{indus5.0_XU2021530}
\bibinfo{author}{Xu, X.}, \bibinfo{author}{Lu, Y.},
  \bibinfo{author}{Vogel-Heuser, B.}, \bibinfo{author}{Wang, L.},
  \bibinfo{year}{2021}.
\newblock \bibinfo{title}{Industry 4.0 and industry 5.0—inception, conception
  and perception}.
\newblock \bibinfo{journal}{Journal of Manufacturing Systems}
  \bibinfo{volume}{61}, \bibinfo{pages}{530--535}.
\newblock \URLprefix
  \url{https://www.sciencedirect.com/science/article/pii/S0278612521002119},
  \DOIprefix\doi{https://doi.org/10.1016/j.jmsy.2021.10.006}.
%Type = Article
\bibitem[{Yeoman et~al.(2022)Yeoman, Weakley, DuBose, Honn, McMurry, Eiter,
  Baker and Poplin}]{62YEOMAN2022103743}
\bibinfo{author}{Yeoman, K.}, \bibinfo{author}{Weakley, A.},
  \bibinfo{author}{DuBose, W.}, \bibinfo{author}{Honn, K.},
  \bibinfo{author}{McMurry, T.}, \bibinfo{author}{Eiter, B.},
  \bibinfo{author}{Baker, B.}, \bibinfo{author}{Poplin, G.},
  \bibinfo{year}{2022}.
\newblock \bibinfo{title}{Effects of heat strain on cognitive function among a
  sample of miners}.
\newblock \bibinfo{journal}{Applied Ergonomics} \bibinfo{volume}{102},
  \bibinfo{pages}{103743}.
\newblock \URLprefix
  \url{https://www.sciencedirect.com/science/article/pii/S0003687022000667},
  \DOIprefix\doi{https://doi.org/10.1016/j.apergo.2022.103743}.
%Type = Article
\bibitem[{Zhu et~al.(2023a)Zhu, Hu, Hu, Wang and Guan}]{60expU_ZHU2023112704}
\bibinfo{author}{Zhu, H.}, \bibinfo{author}{Hu, M.}, \bibinfo{author}{Hu, S.},
  \bibinfo{author}{Wang, H.}, \bibinfo{author}{Guan, J.},
  \bibinfo{year}{2023}a.
\newblock \bibinfo{title}{Effects of hot-humid exposure on human cognitive
  performance under sustained multi-tasks}.
\newblock \bibinfo{journal}{Energy and Buildings} \bibinfo{volume}{279},
  \bibinfo{pages}{112704}.
\newblock \URLprefix
  \url{https://www.sciencedirect.com/science/article/pii/S0378778822008751},
  \DOIprefix\doi{https://doi.org/10.1016/j.enbuild.2022.112704}.
%Type = Article
\bibitem[{Zhu et~al.(2023b)Zhu, Wang, Hu, Ma, Su and Wang}]{zhu2023cognitive}
\bibinfo{author}{Zhu, H.}, \bibinfo{author}{Wang, Y.}, \bibinfo{author}{Hu,
  S.}, \bibinfo{author}{Ma, L.}, \bibinfo{author}{Su, H.},
  \bibinfo{author}{Wang, J.}, \bibinfo{year}{2023}b.
\newblock \bibinfo{title}{Cognitive performances under hot-humid exposure: An
  evaluation with heart rate variability}.
\newblock \bibinfo{journal}{Building and Environment} \bibinfo{volume}{238},
  \bibinfo{pages}{110325}.
\newblock \URLprefix
  \url{https://www.sciencedirect.com/science/article/pii/S0360132323003529},
  \DOIprefix\doi{https://doi.org/10.1016/j.buildenv.2023.110325}.

\end{thebibliography}
